\documentclass[%
11pt,
onecolumn,
tightenlines,
superscriptaddress,
preprintnumbers,
nofootinbib,
amsmath,amssymb,amsthm,
physrev,
eqsecnum,tikz,
]{revtex4-2}

\usepackage{isomath}
\usepackage{amsmath,amsthm}
\usepackage{amsbsy}
\usepackage{amssymb}
\usepackage{amscd}
\usepackage{amsfonts}
\usepackage{stmaryrd}
\usepackage{siunitx}
\usepackage{euscript}
\usepackage[utf8]{inputenc}
\usepackage[T1]{fontenc}
\usepackage{newtxtext} 
\everymath{\displaystyle}
\usepackage{exscale}

\usepackage{graphicx}
\usepackage{boxedminipage}
\usepackage{calc}
\usepackage[usenames,dvipsnames]{xcolor}
\graphicspath{ {media/} }
\usepackage[caption=false,justification=centerlast]{subfig}

\usepackage{setspace}
\usepackage{enumitem}
\setitemize{noitemsep,topsep=0pt,parsep=0pt,partopsep=0pt}
\setenumerate{noitemsep,topsep=0pt,parsep=0pt,partopsep=0pt}
\setdescription{noitemsep,topsep=0pt,parsep=0pt,partopsep=0pt}

\usepackage[colorinlistoftodos, color=green!40,prependcaption]{todonotes}

\usepackage{soul} % highlighting

\usepackage[,colorlinks,urlcolor=blue,citecolor=black]{hyperref}
\usepackage[normalem]{ulem}

\usepackage[small]{titlesec}

\titlespacing*{\section}{0pt}{12pt plus 4pt minus 2pt}{2pt plus 2pt minus 2pt}
\titlespacing*{\subsection}{0pt}{12pt plus 4pt minus 2pt}{2pt plus 2pt minus 2pt}
\titlespacing*\subsubsection{0pt}{12pt plus 4pt minus 2pt}{2pt plus 2pt minus 2pt}
\titlespacing*\paragraph{0pt}{12pt plus 4pt minus 2pt}{2pt plus 2pt minus 2pt}

\makeatletter

    \renewcommand*{\p@subsection}{}
    
    \renewcommand*{\p@subsubsection}{}
\makeatother

\setuptodonotes{inline}
\usepackage{latexsym}
\usepackage{floatflt,graphicx}
\usepackage{amssymb}
\usepackage{amsmath}
\usepackage[plain]{fancyref}
\usepackage{xcolor}
\usepackage{lineno}
\usepackage[colorinlistoftodos, color=green!40,prependcaption]{todonotes}
\usepackage{verbatim}
\usepackage{mathtools}

\newcommand*{\fancyrefapplabelprefix}{app}
\frefformat{plain}{\fancyrefapplabelprefix}{appendix~#1}
\Frefformat{plain}{\fancyrefapplabelprefix}{Appendix~#1}
\frefformat{main}{\fancyrefapplabelprefix}{appendix~#1}
\Frefformat{main}{\fancyrefapplabelprefix}{Appendix~#1}

\usepackage{isomath}
\usepackage{amsmath}
\usepackage{amssymb}
\usepackage{amscd}
\usepackage{amsfonts}

\newcommand{\bfn}{{\mathbold n}}

\newcommand{\bfr}{{\mathbold r}}

\newcommand{\bfE}{{\mathbold E}}
\newcommand{\bfF}{{\mathbold F}}

\newcommand{\bfP}{{\mathbold P}}

\graphicspath{ {figs/} }

\DeclareMathOperator{\erfc}{erfc}
\DeclareMathOperator{\erfi}{erfi}
\DeclareMathOperator{\erf}{erf}

\newcommand{\probSym}{\pi}
\newcommand{\prob}[1]{\probSym\left(#1\right)}
\newcommand{\tranProb}[2]{p\left(#1 \rightarrow #2\right)}
\newcommand{\accProb}[2]{a\left(#1 \rightarrow #2\right)}
\newcommand{\trialProb}[2]{t\left(#1 \rightarrow #2\right)}
\newcommand{\pSpace}{\Gamma}
\newcommand{\df}[1]{\mathrm{d}#1}
\newcommand{\weightSym}{w}
\newcommand{\weight}[1]{\weightSym\left(#1\right)}
\newcommand{\avg}[1]{\left\langle#1\right\rangle}
\newcommand{\modAvg}[1]{\avg{#1}_{\probSym}}
\newcommand{\solidAngle}{\Omega}
\newcommand{\EDimless}{\tilde{E}}
\newcommand{\mStateX}{\boldsymbol{\varphi}}
\newcommand{\mStateY}{\boldsymbol{\upsilon}}
\newcommand{\unitsphereN}[1]{\mathbb{S}^{#1}}
\newcommand{\unitsphere}{\unitsphereN{2}}
\newcommand{\totalPot}{\mathcal{V}}

\renewcommand{\hl}[1]{#1}
\newenvironment{hlbreakable}%
{}%
{}

\begin{document}

\preprint{To appear in Journal of the Mechanics and Physics of Solids (doi: \href{https://doi.org/10.1016/j.jmps.2021.104658}{10.1016/j.jmps.2021.104658})}

\title{\Large{Statistical mechanics of a dielectric polymer chain \hl{in the force ensemble}}}

\author{Matthew Grasinger}
    \email{matthew.grasinger.ctr@afresearchlab.com}
    \affiliation{Materials and Manufacturing Directorate, Air Force Research Laboratory}
    \affiliation{UES, Inc.}

\author{Kaushik Dayal}
    \affiliation{Department of Civil and Environmental Engineering, Carnegie Mellon University}
    \affiliation{Department of Materials Science and Engineering, Carnegie Mellon University}
    \affiliation{Center for Nonlinear Analysis, Department of Mathematical Sciences, Carnegie Mellon University}

\author{Gal deBotton}
    \affiliation{Department of Mechanical Engineering, Ben-Gurion University}
    \affiliation{Department of Biomedical Engineering, Ben-Gurion University}

\author{Prashant K. Purohit}
    \email{purohit@seas.upenn.edu}
    \affiliation{Department of Mechanical Engineering and Applied Mechanics, University of Pennsylvania}

\date{\today}

%%%%%%%%%%%%%%%%%%%%%
%%%%%%%%%%%%%%%%%%%%%
%%%%%%%%%%%%%%%%%%%%%
%%%%%%%%%%%%%%%%%%%%%

\begin{abstract}
    Constitutive modeling of dielectric elastomers has been of long standing
    interest in mechanics. Over the last two decades rigorous constitutive models 
    have been developed that couple the 
    electrical response of these polymers with large deformations characteristic 
    of soft solids. A drawback of these models is that unlike classic models of 
    rubber elasticity they do not consider the coupled electromechanical response 
    of single polymer chains which must be treated using statistical mechanics. 
    The objective of this paper is to compute the stretch and polarization of 
    single polymer chains subject to a fixed force and fixed electric field using
    statistical mechanics. We assume that the dipoles induced by the applied
    electric field at each link do not interact with each other and compute the 
    partition function using standard techniques. We then calculate the stretch 
    and polarization by taking appropriate derivatives of the partition function 
    and obtain analytical results in various limits. We also perform Markov chain Monte Carlo 
    simulations using the Metropolis and umbrella sampling methods, as well as develop a new sampling method which improves convergence by exploiting a symmetry inherent in dielectric polymer chains.
    The analytical expressions are shown to agree with the Monte Carlo results over a
    range of forces and electric fields. Our results complement recent 
    work on the statistical mechanics of electro-responsive chains which obtains 
    analytical expressions in a different ensemble.
\end{abstract}

\maketitle

\section{Introduction} \label{sec:intro}
There has been a paradigm shift in recent years towards drawing inspiration from, and integrating robotics and electronics more seamlessly with natural and biological systems.
This has lead to a great effort toward discovering and developing soft materials and associated mechanisms which couple (potentially large) deformation to electromagnetic fields, light, and other fields which can also be readily and rapidly controlled
~\cite{ware2016localized,majidi2014soft,o2008review,chen2020design,chen2021interplay,shian2015dielectric,brochu2012dielectric,lu2014flexible,bauer201425th,bartlett2016stretchable,bar-cohen2001electroactive,carpi2011electroactive,castaneda2011homogenization,galipeau2013finite,erol2019microstructure,huang2012giant,kim2007electroactive,lopez2014elastic,araromi2014rollable,lau2017dielectric,pourazadi2019investigation,zhao2019mechanics,zadan2021liquid,liao2021soft,zolfaghari2020network,babaei2021torque,zhao2014harnessing,overvelde2015amplifying,keplinger2012harnessing,grasinger2020architected,grasingerIPflexoelectricity,cohen2016electromechanical,tutcuoglu2014energy,kim2019ferromagnetic,kim2018printing,xu2013stretchable,xu2014soft,rogers2010materials,kang2016bioresorbable,kwak2020wireless,gdbetal07mams,tianetal12jmps,rudyetal12ijnm}.
Dielectric elastomers (DEs)--soft materials which polarize and deform in response to an applied electric field--in particular, have emerged as excellent candidates for soft sensors, soft robotics~\cite{chen2020design,chen2021interplay,shian2015dielectric,brochu2012dielectric,lu2014flexible,bauer201425th,araromi2014rollable,lau2017dielectric,pourazadi2019investigation,rogers2010materials,duduta2019realizing}, and biomedical devices such as Braille displays, deformable lenses and haptic devices~\cite{o2008review}.

Earlier dielectric elastomer actuation mechanisms consisted primarily of a thin film sandwiched between compliant electrodes.
A voltage difference is applied across the electrodes, the DE polarizes and compresses across its thickness and, because of the Poisson effect, expands in the plane of the electrodes~\cite{pelrine2000high,pelrine2001dielectric,kofod2008static,kofod2005silicone,kofod2003actuation,kollosche2012complex,wissler2007mechanical}.
However, recently, new types of mechanisms have been developed.
For example, in~\citet{grasingerIPtorque}, it was shown that applying an initial shear deformation and appropriate constraints leads to a shear-mode of electromechanical actuation; and, in~\citet{hajiesmaili2019reconfigurable}, the authors varied electrode geometries through multiple layers of DEs in order to create spatially varying electric fields and realize shape-morphing.
Similar types of mechanisms--namely, introducing (possibly heterogeneous) inital stresses, heterogeneous (elastic) material properties, and/or heterogeneous electric fields--have even been used to develop dielectric elastomer-based robotic grippers~\cite{araromi2014rollable,lau2017dielectric,pourazadi2019investigation,shian2015dielectric},
shape-morphing fibers~\cite{chortos2021printing},
and shape-morphing circular sheets~\cite{hajiesmaili2019voltage}.
These recently developed electromechanical mechanisms show the excellent potential of dielectric elastomers for achieving biomimetic actuation and soft actuation with a large number of degrees of freedom.

However, despite these many advantages, dielectric elastomers are limited by a relatively weak electromechanical coupling such that large electric fields, and therefore large voltages and power sources, are often required to achieve meaningful actuation forces or deformations.
To address the challenge presented by weak couplings in soft multifunctional materials, there has been much development in composite elastomers~\cite{castaneda2011homogenization,galipeau2013finite,siboni2014fiber,liao2021soft,zadan2021liquid,zolfaghari2020network,gdbetal07mams,tianetal12jmps}, architected elastomer networks~\cite{grasinger2020architected,grasingerIPflexoelectricity}, and in harnessing geometric instabilities~\cite{chen2021interplay,babaei2021torque,zhao2014harnessing,overvelde2015amplifying,keplinger2012harnessing,rudyetal12ijnm}.
The latter technique, namely harnessing geometric instabilities, suggests another kind of mechanism for enhancing electromechanical coupling in DEs: in contrast to instabilities at the macroscale, one may inquire whether instabilities at the molecular scale (i.e. phase transitions) may also be leveraged.
Towards this end, it is the goal of this paper to contribute a rigorous and in-depth study of the thermodynamics and statistical mechanics of dielectric polymer chains.
This represents a first step toward understanding possible phase transitions in DE chains and, in confluence with the existing body of literature on the subject~\cite{cohen2016electroelasticity,grasinger2020statistical,itskov2018electroelasticity}, can help inform the design of composite materials and architected elastomer networks.

Although past work has had success in understanding and modelling the electromechanical behavior of dielectric elastomers through continuum-scale constitutive models (based on considerations of symmetry and strain invariants)~\cite{huang2012giant,zhang2017nonlinear,wissler2007mechanical,kofod2008static,kofod2005silicone,kofod2003actuation,kollosche2012complex,pelrine2000high,pelrine2001dielectric,tutcuoglu2014energy,he2009dielectric,bozlar2012dielectric,zhao2007method,zhang2016method,li2016voltage,zhao2010theory,zurlo2017catastrophic,henann2013modeling}, such approaches cannot capture the role of polymer chains, their orientations, and their connectivity within a broader DE network.
As such, they often miss electrically-induced stresses and deformations that occur within the network as chains deform to align their dipoles with the local electric field~\cite{cohen2016electroelasticity,cohen2016electromechanical,grasinger2020statistical,grasinger2020architected,grasingerIPtorque,itskov2018electroelasticity}.
Instead, in recognition of the ground-breaking role of statistical mechanics and polymer network modelling in understanding classical rubber elasticity~\cite{treloar1975physics,kuhn1942beziehungen,weiner2012statistical,arruda1993threee,boyce2000constitutive,flory1944network,miehe2004micro,james1943theory}, recent work has aimed to provide statistical-mechanics based approaches to the modelling of dielectric polymer chain-scale phenomena and its impact on the electroelasticity of DEs~\cite{cohen2016electroelasticity,cohen2016electromechanical,grasinger2020statistical}.
The first such works appear to be~\citet{cohen2016electroelasticity} and~\citet{cohen2016electromechanical} which considered a fixed end-to-end vector (i.e. the vector from the beginning of chain to the end of the chain) and used statistical-mechanical principles to derive chain-scale couplings between the local electric field, net dipole of the chain, and deformation of the chain\footnote
{
    An important body of work is the statistical mechanics of polyelectrolytes, i.e. 
    charged bio-macromolecules typically in solution \cite{argudo2012dependence,wang2004self,shen2017electrostatic}.
    There is a key difference between polyelectrolytes and dielectric elastomers: the former typically consist of fixed (i.e., independent of electric field) charges carried on monomers; the latter consist of field-responsive dipoles carried on monomers, with the dipole orientation coupled to the monomer orientation.
}.
The statistical mechanics formulation in the fixed end-to-end vector ensemble presents a formidable challenge; hence, the authors assumed smallness of parameters and used Taylor expansion approximations to obtain the density of link (or monomer) orientations within a chain.
\citet{grasinger2020statistical} built upon this work by considering various regimes of electric field magnitudes and chain stretches.
Then, by using what was known about the limiting behavior, a free energy approximation was constructed which agreed well with numerical (mean-field theory) solutions~\cite{grasinger2020statistical}.

There are, however, still gaps that remain in the literature on the statistical mechanics of dielectric polymer chains.
First, we remark that both the closed-form approximations and numerical solutions obtained in past work are specific to the fixed end-to-end vector ensemble and utilize mean-field theory (alternatively, referred to as the maximum-term approach~\cite{hill1986statistical})~\cite{cohen2016electroelasticity,cohen2016electromechanical,grasinger2020statistical,itskov2018electroelasticity}.
However, only considering one ensemble is unlikely to provide a complete description of a thermodynamic system--as it is well known that different ensembles lead to different results for finite-sized systems~\cite{gross1996microcanonical,deserno1997tricriticality,kastner1999magnetic,weiner2012statistical,tadmor2011modeling}, and there is no guarantee that any two ensembles are equivalent even for arbitrarily large systems (i.e. in the ``thermodynamic limit'')~\cite{touchette2004introduction}.
\hl{For instance, it is well known that there are differences in behavior between the end-to-end vector and force ensembles for the elasticity of freely jointed polymer chains~\cite{weiner2012statistical,treloar1975physics}; and the differences only become negligible in the long chain limit~\cite{treloar1975physics}.}
Further, for a mean-field theory to be an accurate representation of the system, we require that the behavior of each individual particle be well-represented by the average behavior over all particles; thus, fluctuations in particle behavior must be small and homogeneous.
It is difficult to justify the assumption of small fluctuations \emph{a priori}, and it can breakdown, even for large systems, at critical points.
Similarly, the differences between various ensembles are greater at critical points in general and phase transitions in some cases~\cite{deserno2004microcanonical,gross1996microcanonical,deserno1997tricriticality,kastner1999magnetic}~\footnote{As a consequence of the non-equivalence of ensembles at critical points, it is often the case that some ensembles are better for studying phase transitions than others~\cite{gross1996microcanonical,deserno1997tricriticality,kastner1999magnetic}.}.
In this work, we study the statistical mechanics of dielectric polymer chains in the fixed force ensemble (more specifically, the chain end-to-end vector is allowed to fluctuate while a fixed force is applied to its end).
There are several benefits to this choice of ensemble:
\begin{enumerate}
  \item The mathematical formulation of the partition function in the fixed force ensemble is easier to evaluate than in the fixed end-to-end vector ensemble.
    As a result, we can evaluate the partition function without resorting to a mean-field theory; and we do so while making less approximations than in past works.
    In a special case, the partition function can even be evaluated exactly.
  \item Markov chain Monte Carlo (MCMC) sampling is also more facile in the fixed force ensemble.
    Therefore we can verify our results against high fidelity MCMC simulations and see that we obtain excellent agreement.
  \item More broadly, this choice of ensemble, and subsequent work, represents a fundamental next step toward a rigorous understanding of the thermodynamics of dielectric polymer chains--including the possibility of exploiting phase transitions as a means of increasing electromechanical actuation.
\end{enumerate}

The remainder of the paper is organized as follows:
\begin{itemize}
  \item In \Fref{sec:kinematics}, we establish the kinematics and energetics of a single dielectric polymer chain.
  \item In \Fref{sec:partition}, we formulate 1) the partition function in the fixed force ensemble, 2) the Gibbs free energy, and 3) the chain stretch and polarization as derivatives of the free energy.
  \item In \Fref{sec:monte}, we outline some of the theoretical and implementation details of the Markov chain Monte Carlo method used in this work--including important modifications which improve the convergence of the algorithm for dielectric polymer chains.
  \item In \Fref{sec:analytical}, we obtain closed-form approximations for various cases: 1) the electric field is large relative to the applied force, 2) the force applied is large relative to the electric field, and 3) a special case where the exact solution may be obtained. The approximations obtained in the various limits are compared with the 
  MCMC results and shown to have excellent agreement.
  \item In \Fref{sec:thermo}, we compare the results obtained using the force ensemble to the results obtained using the end-to-end vector ensemble with an emphasis on important phenomena that the two approaches share, implications for the thermodynamic limit, and implications for dielectric elastomer networks.
  \item In \Fref{sec:conclusions}, the main contributions of this work and some of its limitations are summarized.
\end{itemize}

\section{Kinematics and energetics of a single chain} \label{sec:kinematics}

\begin{figure}
\centering
\includegraphics{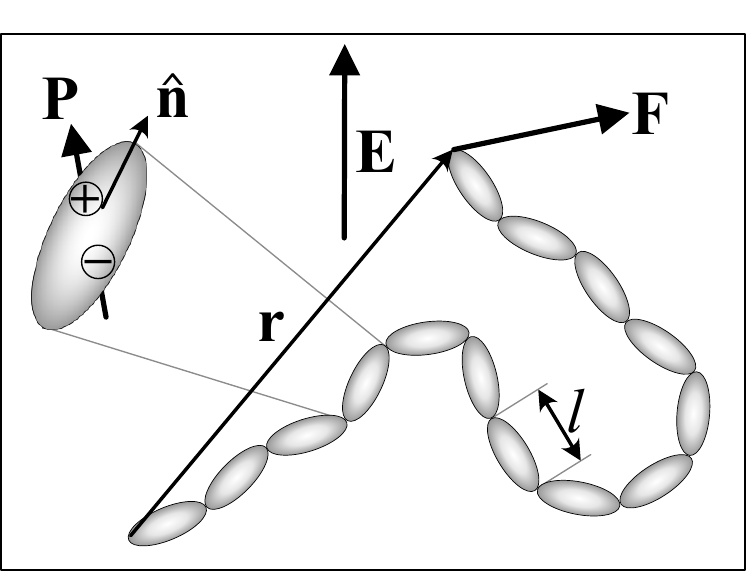}
\caption{A sketch of a chain and the force $\mathbf{F}$ and the electric field $\mathbf{E}$ acting upon it.}
\label{fig:dechain}
\end{figure}

We assume that our chain consists of $N$ rigid links. Each link is of length $l$ and
is free to be oriented in any direction in 3D space (Fig.~\ref{fig:dechain}). There is no energy 
penalty associated with the angle between successive links or with excluded 
volume interactions as in the freely-jointed-chain (FJC) model of polymer 
elasticity. The unit vector along a link is
\begin{equation}
 \hat{\mathbf{n}} = \sin\theta \cos\phi \mathbf{e}_{x} + \sin\theta
 \sin\phi \mathbf{e}_{y} + \cos\theta \mathbf{e}_{z},
\end{equation}
where $\theta$ is the angle between the link and the $z$-axis and $\phi$ is the
angle made by projection of the link on the xy-plane with the $x$-axis. 
$[\mathbf{e}_{x} \quad \mathbf{e}_{y} \quad \mathbf{e}_{z}]$ are unit vectors 
of a Cartesian coordinate system. One
may recall that in the freely-jointed-chain model the $z$-direction is chosen
to align with the force $\mathbf{F}$ acting on the chain, so 
$\mathbf{F} = F\mathbf{e}_{z}$\footnote{$\mathbf{F}$  should not be confused with the deformation gradient tensor of continuum mechanics since we do not use it anywhere in this paper which treats the mechanics of single chain of dielectric polymer.}. We have a given electric field
vector $\mathbf{E}$ in addition to the force vector $\mathbf{F}$, so our 
choice of $z$-axis will be dictated by convenience of calculation of the 
partition function in various limits. In all our calculations one end of the
chain will be fixed to the origin and the position in 3D space of the other
end of the chain will be calculated. 

Next, we consider the energetics of the chain. Each link is dielectric, so its dipole moment vector is given by
\begin{equation}
  \mathbf{P} = \epsilon_0 \mathbf{K}\mathbf{E} = \epsilon_0 \left[K_{1}\left(\hat{\mathbf{n}} \otimes 
      \hat{\mathbf{n}}\right) + K_{2} \left(\mathbf{I} - \hat{\mathbf{n}}\otimes 
  \hat{\mathbf{n}}\right)\right]\mathbf{E},
\end{equation}
where $\epsilon_0$ is the permittivity of free space, and $K_{1}$ and $K_{2}$ are given constants.
Let us take $\mathbf{E} = E_{x}\mathbf{e}_{x} + E_{y}\mathbf{e}_{y} 
 + E_{z}\mathbf{e}_{z}$ fixed. Then, the energy of the link due to the 
induced dipole caused by the electric field is $V_{elec} = \frac{\epsilon_0}{2}
\mathbf{E}\cdot \mathbf{K}\mathbf{E} - \mathbf{P}\cdot\mathbf{E} = -\frac{1}{2}
\mathbf{P}\cdot \mathbf{E}$, or
\begin{equation} \label{eq:Velec}
 V_{elec} =-\frac{1}{2}\mathbf{P}\cdot\mathbf{E} = - \frac{\epsilon_0\left(K_{1} - K_{2}\right)}{2}
 \left(E_{x}\sin\theta\cos\phi + E_{y}\sin\theta\sin\phi + E_{z}\cos\theta\right)^{2}
 - \epsilon_0 \frac{K_{2}}{2}\left(E_{x}^{2} + E_{y}^{2} + E_{z}^{2}\right).
\end{equation}
For ease of calculation and because of its validity for most practical cases of interest, we make the fairly standard assumption that the energy of dipole-dipole interactions are much less than the thermal energy and can therefore be neglected~\cite{grasinger2020statistical,cohen2016electroelasticity}, i.e. 
$\epsilon_0 \left(E \times \max \left\{K_1, K_2\right\} \right)^2 / l^3 \ll k_B T$.
Having made the assumption of negligible dipole-dipole interactions precise, moving forward, we set $\epsilon_0 = 1$ with the understanding that $K_1 \rightarrow \epsilon_0 K_1$ and $K_2 \rightarrow \epsilon_0 K_2$ recovers the equivalent formulation in which $\epsilon_0$ is included explicitly\footnote
{
We highlight that treating the dipole-dipole interactions generally leads to a challenging nonlocal problem that has typically only been treated at zero temperature \cite{james1994internal,marshall2014atomistic}.
}.

%%The second term in $V_{elec}$ will get factored out of the partition sum
%%because it is a constant independent of the angles. Recall that the partition
%%sum will ultimately be an integral over all values of the angles $\theta$ 
%%and $\phi$. 
The potential energy of the applied force 
$\mathbf{F} = F_{x}\mathbf{e}_{x} + F_{y}\mathbf{e}_{y}
 + F_{z}\mathbf{e}_{z}$ is 
\begin{equation}
 V_{mech} = -F_{x}l\sin\theta\cos\phi - F_{y}l\sin\theta\sin\phi 
 - F_{z}l\cos\theta,
\end{equation} 
where $l$ is the length of the link. The total energy of a link whose unit vector 
is given by the angles $\left(\theta,\phi\right)$ is
\begin{equation}
  V\left(\theta,\phi\right) = V_{elec} + V_{mech}.
\end{equation}

\section{Partition function of the chain} \label{sec:partition}
The partition function for a chain of $N$ non-interacting links is 
$Z = Z_{1}^{N}$ where $Z_{1}$ is the partition function for one link and is 
given by:
\begin{equation} \label{eq:Z1parti}
 Z_{1} = \int_{0}^{2\pi} \int_{0}^{\pi}\exp\left(
 -\frac{V\left(\theta,\phi\right)}{k_{B}T}\right)\sin\theta \: \df{\theta} \df{\phi}.
\end{equation}
When $\mathbf{E} = 0$ and $\mathbf{F} = F\mathbf{e}_{z}$ this integral leads to
the well-known result 
$Z_{1} = 4\pi\frac{k_{B}T}{F_{z}l}\sinh\left(\frac{F_{z}l}{k_{B}T}\right)$ for the 
partition function of one link of a freely jointed chain. Then, the free energy of the 
chain is
\begin{equation} 
  G\left(\mathbf{F},\mathbf{E},T\right) = -Nk_{B}T\log Z_{1}
\end{equation}
The average position of the other end of the chain is given by 
$\mathbf{r} = r_{x}\mathbf{e}_{x} + r_{y}\mathbf{e}_{y} + r_{z}\mathbf{e}_{z}$
where
\begin{equation}
 r_{x} = -\frac{\partial G}{\partial F_{x}}, \qquad 
 r_{y} = -\frac{\partial G}{\partial F_{y}}, \qquad 
 r_{z} = -\frac{\partial G}{\partial F_{z}}.
\end{equation}
The stretches may then be evaluated as $\lambda_{x} = \frac{r_{x}}{Nl}$,
$\lambda_{y} = \frac{r_{y}}{Nl}$ and $\lambda_{z} = \frac{r_{z}}{Nl}$.
One may recall that the average end-to-end distance of the chain in the FJC
model in which $\mathbf{F} = F\mathbf{e}_{z}$ is along the $z$-direction 
is given by 
\begin{equation}
 \langle z \rangle = -\frac{dG}{dF} = \frac{Nk_{B}T}{Z_{1}}
 \frac{\partial Z_{1}}{\partial F}, 
\end{equation}
so that the stretch of the chain is
\begin{equation}
 \lambda = \frac{\langle z \rangle}{Nl} = \frac{k_{B}T}{lZ_{1}}
 \frac{\partial Z_{1}}{\partial F}. 
\end{equation}
The average polarizations are given by
\begin{equation} \label{eq:polarization}
 P_{x} = -\frac{\partial G}{\partial E_{x}}, \qquad 
 P_{y} = -\frac{\partial G}{\partial E_{y}}, \qquad 
 P_{z} = -\frac{\partial G}{\partial E_{z}}. 
\end{equation}
All of these expressions require us to compute $Z_{1}$ which is the
partition function of one link. This computation is in general quite
involved, however, we can make analytical progress in some special cases 
which will be described in detail later. Ensemble averages of stretch and polarization for a chain may also be 
computed using Monte Carlo simulations. We describe both these methods below
and show that they produce identical results in various regimes.

\section{Monte Carlo methods for dielectric polymer chains} \label{sec:monte}

This section outlines the theoretical and implementation details related to the Markov chain Monte Carlo method used in this work--with an emphasis on important modifications that were made to the standard treatment in order to improve its convergence and address certain challenges specific to dielectric polymer chains.
%A broader overview of the key concepts of Monte Carlo methods in statistical mechanics can be found in~\citet{krauth2006statistical}.
Results from the simulations are provided and discussed in later sections.

%It is of course well known that using numerical quadrature is generally impractical for evaluating the partition function in statistical mechanics.
%This is because evaluation of the partition function involves integration in $3N$ (or $6N$) dimensional space.
%Doubling the density of samples of the quadrature in such a high dimensional space means increasing the number of samples by $2^{3 N}$.
%Since $N \sim 100$--$10000$, this is inconceivable.
%To deal with this \emph{``curse of dimensionality''}, we resort to stochastic integration.
%An overview of the key concepts of Monte Carlo methods in statistical mechanics can be found in~\citet{krauth2006statistical}.
%Below, we only introduce the details most relevant to this work.

\hl{Monte Carlo methods are amenable to approximating solutions to problems that allow a probabilistic interpretation.}
%Thus, while we can not, and do not, aim to obtain the free energy via Monte Carlo simulation, it is useful for approximating the thermodynamic state variables of a system.
Thus, Monte Carlo methods are effective for simulating the properties of a thermodynamic system because state variables (e.g. in the case of an ideal gas: temperature, pressure, volume, etc.; in the present work: end-to-end vector, net chain dipole, etc.) correspond to averages over thermal fluctuations.
The method works by approximating these averages through random sampling of the appropriate probability distribution.
Direct random sampling is difficult if the domain of integration is unbounded; and it is inefficient if the probability distribution consists of a sharp peak(s) and is decidedly nonuniform.
Markov chain sampling is an effective way to mitigate these challenges.
In a Markov chain, one constructs a chain of sampled configurations~\footnote{In statistical mechanics, configurations of the system are often called microstates.} where, starting with an initial configuration, a trial for each new configuration is proposed by perturbing the previous configuration by some small amount.
Whether or not the trial is accepted is based on a chosen \emph{acceptance criteria} which helps model the underlying probability distribution that is being sampled.
To truly model the underlying probability distribution, we require that it must sample configurations $\mStateX \in \pSpace$ with probabilities $\prob{\mStateX}$, where $\mStateX$ is a configuration within the (possibly uncountable) collection of admissible configurations, denoted by $\pSpace$, and where $\prob{\mStateX}$ denotes the probability of the system being in configuration $\mStateX$.
For dielectric polymer chains, $\pSpace = \unitsphere \times \unitsphere \times ... \times \unitsphere = \unitsphereN{2N}$ and $\mStateX = \left(\hat{\mathbf{n}}_1, ..., \hat{\mathbf{n}}_N\right)$; that is, $\mStateX$ specifies the directions of each of the links in the chain.
In the ensemble of interest in this work, $\prob{\mStateX} \propto \exp \left(-\totalPot\left(\mStateX\right) / k_B T\right) \solidAngle\left(\mStateX\right)$ where $\totalPot\left(\mStateX\right)$ is the total potential energy of the chain and $\solidAngle\left(\mStateX\right)$ is the $2N$ dimensional solid angle of $\mStateX$; i.e.,
\begin{equation}
    \begin{split}
        \totalPot\left(\mStateX\right) &= \sum_{i=1}^N V\left(\theta_i, \phi_i\right),\\
        \solidAngle\left(\mStateX\right) &= \prod_{i=1}^N \sin \theta_i \, \df{\theta_i} \df{\phi_i}.
    \end{split}
\end{equation}

Consider another possible configuration of the system $\mStateY \in \pSpace$.
Let $\tranProb{\mStateX}{\mStateY}$ denote the transition probability for moving from configuration $\mStateX$ to configuration $\mStateY$.
%Assume $\pSpace$ is continuous (i.e. not discrete).
Then, because the system must transition to some state (not excluding the current state), normalization requires that
\begin{equation} \label{eq:dt-1}
  \int_{\pSpace} \tranProb{\mStateX}{\mStateY} \, \df{\mStateY} = 1, \quad \forall \mStateX \in \pSpace.
\end{equation}
Also, by definition
\begin{equation} \label{eq:dt-2}
  \prob{\mStateX} = \int_{\pSpace} \prob{\mStateY} \tranProb{\mStateY}{\mStateX} \, \df{\mStateY}.
\end{equation}
Multiplying \eqref{eq:dt-1} by $\prob{\mStateX}$ and substituting into \eqref{eq:dt-2}, we arrive at:
\begin{equation} \label{eq:dt-3}
  \int_{\pSpace} \prob{\mStateX} \tranProb{\mStateX}{\mStateY} \, \df{\mStateY} = \int_{\pSpace} \prob{\mStateY} \tranProb{\mStateY}{\mStateX} \, \df{\mStateY},
\end{equation}
which we require to hold for all possible pairs of configurations $\mStateX$ and $\mStateY$.
In general, depending on the specific features of $\pSpace$, there may be more than one way to satisfy \eqref{eq:dt-3}.
However, one can easily see that a simple condition, called \emph{detailed balance}~\cite{krauth2006statistical}:
\begin{equation}
  \prob{\mStateX} \tranProb{\mStateX}{\mStateY} = \prob{\mStateY} \tranProb{\mStateY}{\mStateX},
\end{equation}
is sufficient for satisfying \eqref{eq:dt-3}.
Nearly every sampling algorithm in the literature is constructed to satisfy detailed balance.

Next, as it will become important in later developments, we decompose the transition probabilities.
Let $\trialProb{\mStateX}{\mStateY}$ and $\accProb{\mStateX}{\mStateY}$ denote the probability of proposing trial move $\mStateX \rightarrow \mStateY$ and accepting the trial move $\mStateX \rightarrow \mStateY$, respectively.
By definition, $\tranProb{\mStateX}{\mStateY} = \trialProb{\mStateX}{\mStateY} \accProb{\mStateX}{\mStateY}$.
For most applications, trial moves are generated isotropically; that is, perturbations are sampled uniformly in every direction.
A standard technique, and one that is employed in the current work, is to choose, at random, one of the links (or in more general language, ``particles'') and perturb it by some random small amount.
To be precise, trial moves are generated by generating a random integer $t \in \left\{1, 2, ..., N\right\}$ and performing the operation
\begin{equation}
  \begin{split}
    \phi_t &\rightarrow \phi_t + \Delta \phi, \\
    \theta_t &\rightarrow \theta_t + \Delta \theta,
  \end{split}
\end{equation}
where $\Delta \phi$ and $\Delta \theta$ are generated from uniform distributions on $\left[-\Delta \phi_{max}, \Delta \phi_{max}\right]$ and $\left[-\Delta \theta_{max}, \Delta \theta_{max}\right]$, respectively; and where $\Delta \phi_{max}$ and $\Delta \theta_{max}$ are determined adaptively to achieve an acceptance ratio of between $0.15$ and $0.55$~\footnote{Specifically, the maximum step sizes are increased if the acceptance ratio is above $0.55$ and decreased if the acceptance ratio is below $0.15$. We chose these thresholds for the acceptance ratio because it is well understood that a proper acceptance ratio ($\sim 0.23$) leads to a better convergence rate~\cite{krauth2006statistical,gelman1996efficient}}.
Because the perturbations of $\phi_t$ and $\theta_t$ are uniformly sampled about $0$, this method for constructing a trial configuration ensures that $\trialProb{\mStateX}{\mStateY} = \trialProb{\mStateY}{\mStateX}$ for all possible pairs $\mStateX$ and $\mStateY$.

Since $\trialProb{\mStateX}{\mStateY} = \trialProb{\mStateY}{\mStateX}$, to satisfy detailed balance, we simply need to choose an acceptance criteria such that $\prob{\mStateX} \accProb{\mStateX}{\mStateY} = \prob{\mStateY} \accProb{\mStateY}{\mStateX}$.
Because of its relative simplicity, its excellent convergence properties, and the fact that it satisfies detailed balance, the most popular acceptance criteria is the  Metropolis-Hastings criteria~\cite{hastings1970monte,metropolis1953equation}:
\begin{equation} %\label{eq:metro-hast}
  \accProb{\mStateX}{\mStateY} = \min \left\{ 1, \frac{\prob{\mStateY}}{\prob{\mStateX}} \right\}.
\end{equation}
In this work, we study the computational statistical mechanics of dielectric polymer chains using the Markov chain Monte Carlo (MCMC) method with the Metropolis-Hastings acceptance criteria~\footnote{The code and precise implementation details used for the simulations in this work can be found at \url{https://github.com/grasingerm/polymer-stats}.}.
All of the MCMC simulations in this paper were run with $10^6$ steps.
For reasons that will become clear in subsequent sections, we  make modifications to the standard MCMC treatment in order to vastly improve its convergence rate for dielectric polymer chains.
One such modification, to the authors' knowledge, is a novel method for improving convergence of dielectric polymer chains and we briefly explain how it can be understood in terms of a similar algorithm that 
exists for the Ising model of ferromagnetism.
As elaborated on later, together, the two techniques suggest a general principle for improving MCMC convergence.

\subsection{Energy barriers in dielectric polymers}

To illustrate why modifications to the standard MCMC treatment are needed, consider the electrical \hl{potential} energy of a link given in \eqref{eq:Velec}.
There is a symmetry such that the energy is invariant with respect to an inversion of the link direction, i.e. invariant with respect to $\hat{\mathbf{n}} \rightarrow -\hat{\mathbf{n}}$.
When $K_2 > K_1$, links attain their minimum \hl{potential} energy when $\mathbf{E} \cdot \hat{\mathbf{n}} = 0$.
In this case, there is a single, continuous energy valley on the unit sphere centered where the plane $\mathbf{E} \cdot \hat{\mathbf{n}} = 0$ cuts through it.
In contrast, however, when $K_1 < K_2$, there are \hl{potential} energy minima which occur in two discrete directions, $\hat{\mathbf{n}} = \hat{\mathbf{E}}$ and $\hat{\mathbf{n}} = -\hat{\mathbf{E}}$.
Importantly, there is an energy barrier between them that scales with $\left|\mathbf{E}\right|^2 \left(K_1 - K_2\right)$.
It is well known that when the \hl{potential} energy landscape is such that there are multiple wells separated by energy barriers, the convergence rate of the standard Metropolis MCMC algorithm is quite slow~\cite{torrie1977nonphysical,kumar1992weighted}.
Thus, when $\left|\mathbf{E}\right|^2 \left(K_1 - K_2\right) \gg k_B T$ and $\left|\mathbf{E}\right|^2 \left(K_1 - K_2\right) \gg \left|\mathbf{F}\right| l$, we find that the convergence rate is poor and that we require modified sampling algorithms.

\subsection{Umbrella sampling and its limitations for dielectric polymers}

The main idea in umbrella sampling is to introduce a pseudopotential in order to reduce or cancel the barriers in the \hl{potential energy} landscape~\cite{torrie1977nonphysical,torrie1977monte,kastner2011umbrella}.
In other words, the probabilities are biased by a weight function which depends on the configuration, $\mStateX$:
\begin{equation}
  \prob{\mStateX} \propto \weight{\mStateX} \exp\left(-\totalPot\left(\mStateX\right) / k_B T\right) \solidAngle\left(\mStateX\right).
\end{equation}
Then the thermodynamic state variables can be obtained by a modified averaging over the MCMC samples:
\begin{equation}
  \avg{A} = \frac{\modAvg{A / \weightSym}}{\modAvg{1 / \weightSym}}.
\end{equation}
where $\avg{A}$ denotes the ensemble average of $A$ and $\modAvg{.}$ denotes the MCMC average using the modified potential.

There are many subtleties related to choosing an appropriate weight function $\weight{\mStateX}$ such that the MCMC sampling is ergodic\footnote{
	\begin{hlbreakable}
	Following standard terminology~\cite{weiner2012statistical,tadmor2011modeling}, we say that a thermodynamic system is ergodic provided ensemble averages (i.e. averages over all of phase space) are equivalent to their corresponding time averages for all phase functions of the system (when its in equilibrium).
	As a consequence, the thermodynamic state of the system is invariant with respect to its initial condition or its time history.
	Similarly, we say that a Markov chain is ergodic provided that its sample averages converge to the phase space average--again, a consequence being that sample averages do not depend on the chain's initial conditions.
	\end{hlbreakable}
} (i.e. does not become trapped in some subset of $\pSpace$) and has a good convergence rate~\cite{frenkel2001understanding}.
In the current work, we introduce the weight function
\begin{equation}
  \weight{\mStateX; \bfF, \bfE} = \begin{cases}
    \exp\left(\eta_1 \left(\eta_2 + \left(1 - \eta_2\right)e^{-\eta_3 |\bfF| l / k_B T}\right)\sum_{i=1}^N \frac{V_{elec}\left(\hat{\mathbf{n}}_i\right)}{k_B T}\right) & K_1 > K_2 \\
    1 & \text{otherwise}
  \end{cases},
\end{equation}
where $\eta_1 \in \left[0, 1\right]$, $\eta_2 \in \left[0, 1\right]$, and $\eta_3 \geq 0$, and find that it works well for a certain range conditions.
Note the following features of the weight function:
\begin{enumerate}
  \item The weight reduces to unity (i.e. there is no biasing) when $K_2 > K_1$.
  \item When the force is small, the weight lifts the \hl{potential} energy wells of $V_{elec}$.
    The magnitude of the lift is determined by the computational parameter $\eta_1$ where $\eta_1 = 0$ corresponds to no lift (i.e. $w = \text{const} = 1$) and $\eta_1 = 1$ corresponds to cancelling the biasing of the electrical energy entirely.
    \item The argument in the exponential of the weight decays exponentially with the magnitude of the applied force (i.e. the weight approaches unity).
    This is desirable since the energy wells are distorted such that they are no longer separated as $\left|\mathbf{F}\right|$ increases.
    The parameters $\eta_2$ and $\eta_3$ determine the fraction of the argument that is retained in the limit of $\left|\bfF\right| \rightarrow \infty$ and the rate of decay, respectively.
    Note that if either $\eta_2 = 1$ or $\eta_3 = 0$ then $w$ does not change with $\left|\mathbf{F}\right|$.
    \item After various trials, values of $\eta_1 = 1.0$, $\eta_2 = 0.25$, and $\eta_3 = 1.0$ were found to be effective for small to moderate electric fields, i.e. $\left|\mathbf{E}\right| \sqrt{\left(K_1 - K_2\right) / k_B T} \leq 3$.
\end{enumerate}
While umbrella sampling proved to be an effective technique for large forces or small to moderate electric fields (i.e. $\left|\mathbf{E}\right| \sqrt{\left(K_1 - K_2\right) / k_B T} \leq 3$), it could no longer obtain good convergence for $K_1 > K_2$, small force, and large field ($\left|\mathbf{E}\right| \sqrt{\left(K_1 - K_2\right) / k_B T} > 3$).
This is because, if one chooses a smaller value of $\eta_1$, the weight function is no longer sufficient to smooth the \hl{potential} energy landscape such that the sampling is effectively ergodic; alternatively, if one chooses $\eta_1 \approx 1$, the Markov chain spends too much time sampling configurations that have negligible probability.
Further, since $\eta_2$ and $\eta_3$ are related to transition of the weight toward $1$ as $\left|\bfF\right| \rightarrow \infty$, adjusting their values do not affect the trade-off between ergodicity and importance. 
In some sense, by biasing against the Boltzmann factor associated with the electrical energy, we negate the entire notion of importance sampling and its benefits.
Instead, we need a sampling technique which is both ergodic and spends the majority of its time sampling higher probability configurations.
Such a sampling technique, which, for dielectric polymer chains, is more elegant and effective than umbrella sampling, is developed in the next subsection.

\subsection{The importance of considering discrete symmetries: flipping links} \label{sec:flipping}
To develop an effective sampling algorithm for dielectric polymer chains, we draw inspiration from the clustering algorithms used to study the ferromagnetic Ising model~\cite{wolff1989collective,swendsen1987nonuniversal}.
The Ising model has a similar challenge in the sense that, in the absence of an applied magnetic field, there are \hl{potential} energy wells which are separated in phase space that consist of all of the spins in the spin-up state or all of the spins in the spin-down state.
When the strength of interactions (often denoted by $J$) is much greater than the thermal energy, the standard sampling technique (of flipping a single spin for each trial move) tends to become stuck near one of the two wells and does not sample phase space ergodically.
Clustering algorithms address this issue by recognizing an important symmetry: taking a large cluster of neighboring spin-up (-down) states and flipping them all to spin-down (-up) results in a negligible change in energy.
Flipping clusters of like states allows the system to be sampled both in a way which respects importance (i.e. spends more time sampling configurations with a lower potential energy) and achieves ergodicity.
What remains to be considered is how clusters are chosen and constructed such that detailed balance is still satisfied.
An overview of these details can be found in \S 5.2.3 of \citet{krauth2006statistical}.

Next, we make an analogy between clusters of spin-up or spin-down states in the Ising model and $\hat{\mathbf{n}}$ for a dielectric monomer.
For dielectric polymer chains, we do not need to consider clusters of monomers, but instead recognize that the symmetry $\hat{\mathbf{n}} \rightarrow -\hat{\mathbf{n}}$ exists for each individual monomer when the electrical \hl{potential} energy is much greater than $\left|\bfF\right| l$.
We therefore modify each trial move by adding an extra step.
After perturbing $\hat{\mathbf{n}_t}$, we flip the monomer (i.e. $\hat{\mathbf{n}_t} \rightarrow -\hat{\mathbf{n}_t}$) with probability $\probSym_{flip}$.
Regardless of the choice for $\probSym_{flip}$, it is still true that $\trialProb{\mStateX}{\mStateY} = \trialProb{\mStateY}{\mStateX}$ (because the operation $\hat{\mathbf{n}_t} \rightarrow -\hat{\mathbf{n}_t}$ is an inversion operation and hence its own inverse); and, therefore, together with the Metropolis-Hastings acceptance criteria, the proposed sampling algorithm still satisfies detailed balance.
However, we do not wish to make a choice near to unity because it will lead to a kind of deterministic periodic sampling that is undesirable, and the choice should not be too low as its effect will be diminished.
Hence, in this work, we make the choice $\probSym_{flip} = 0.5$.
Together with the umbrella sampling technique outlined in the previous section~\footnote{While the results presented herein make use of both techniques, namely umbrella sampling and link flipping, we remark that preliminary results suggest that employing link flipping alone without umbrella sampling out performs employing umbrella sampling alone (without link flipping).
A more precise comparison between the two sampling strategies, as well as a more general investigation of exploiting discrete symmetries in Markov chain Monte Carlo simulations, will be addressed in future work.}, we find our MCMC simulations show excellent convergence for most electric fields, applied forces, etc.
The only exceptional cases were when $K_1 > K_2$, the electric field was large, the applied force was large, and the electric field and force were orthogonal, or nearly orthogonal, to each other.
Some reasons as to why these particular cases were still difficult to sample, and its relation to possible phase transitions in dielectric polymer chains, are discussed further in \Fref{sec:special}.

Before closing this section, it is worthwhile to frame the principles discussed above in relation to computational statistical physics more broadly.
The clustering algorithms for the Ising model and the flipping of link directions in the current work together suggest a more general principle: \emph{discrete symmetries often lead to \hl{potential} energy landscapes with separate, symmetric wells, separated by large barriers; thus, a proper sampling algorithm should aim to respect discrete symmetries as a means to achieve both importance sampling and ergodicity}.
Interestingly, both the symmetry in the Ising model and--as we will show in \Fref{sec:special}--the discrete symmetry in dielectric polymer chains are directly related to second-order phase transitions.

\hl{We emphasize, however, that the free energy appears to be convex, as we expect from the outcome of statistical mechanics when the ensemble is not further restricted \cite{tadmor2011modeling,weiner2012statistical}.
There is evidence of this in past work in the fixed end-to-end vector ensemble~\cite{grasinger2020statistical}, as well as in the present work in the fixed force ensemble.
}

\section{Analytical computation of the partition function} \label{sec:analytical}
In this section we will compute the partition function and free energy
of the chains analytically. We will treat separately several cases that each
corresponds to a different limit and/or choice of coordinate system. The chain behavior and free energy are of course independent of the choice of coordinate system. We will find that it depends on the invariants 
$\mathbf{E}\cdot\mathbf{E}$, $\mathbf{E}\cdot\mathbf{F}$ and 
$\mathbf{F}\cdot\mathbf{F}$.

\begin{hlbreakable}
The analytical results will then be compared to the Monte Carlo simulations for $N = 100$ and various applied forces and electric fields.
We remark that while it is apparent from \eqref{eq:Z1parti} that the $N$-dependence of thermodynamic quantities in the force ensemble is trivial ($\lambda$ is independent of $N$ and $\bfP$ scales linearly with $N$), the dependence on $N$ is nontrivial in the fixed end-to-end vector ensemble~\cite{treloar1975physics,weiner2012statistical,grasinger2020statistical}.
The choice of $N$ is made for the purpose of directly comparing the results of this section with the electroelastic behavior of dielectric chains in the fixed end-to-end vector ensemble.
The force and end-to-end vector ensembles are compared in \fref{sec:thermo}.
\end{hlbreakable}
 
\subsection{Case 1: Large electric field}
Without loss of generality, when the electric field is large it is convenient to assume that the $z$-axis is aligned with the applied electric field. 
So, $\mathbf{E} = E_{z}\mathbf{e}_{z}$. Then, the total energy of a link takes the form:
\begin{equation}
 V(\theta,\phi) = V_{elec} + V_{mech} 
 = -\frac{K_{1} - K_{2}}{2} E_{z}^{2}\cos^{2}\theta
 - \frac{K_{2}}{2}E_{z}^{2} - F_{x}l\sin\theta\cos\phi 
 - F_{y}l\sin\theta\sin\phi - F_{z}l\cos\theta.
\end{equation}   
Again, the partition function we want to compute is
\begin{equation}
 Z_{1} = \int_{0}^{2\pi}\int_{0}^{\pi}
 \exp\left(-\frac{V(\theta,\phi)}{k_{B}T}\right)\sin\theta \: \df{\theta} \df{\phi}
\end{equation}
Let us do the $d\phi$ integral first by focusing on just the $\phi$ 
dependent terms above. The integral is:
\begin{equation}
 \int_{0}^{2\pi}\exp\left(\frac{F_{x}l\sin\theta\cos\phi + F_{y}l\sin\theta
 \sin\phi}{k_{B}T}\right) \: \df{\phi} = 2\pi I_{0}\left(\frac{l\sin\theta}{k_{B}T}
 \sqrt{F_{x}^{2} + F_{y}^{2}}\right),
\end{equation}
where $I_{0}(z)$ is a modified Bessel function of the first kind. The
partition function then becomes:
\begin{equation} \label{eq:blue}
 Z_{1} = 2\pi\int_{0}^{\pi}\exp\left(-\frac{-\frac{K_{1} - K_{2}}{2} E_{z}^{2}
 \cos^{2}\theta - \frac{K_{2}}{2}E_{z}^{2} - F_{z}l\cos\theta}{k_{B}T}\right)
 I_{0}\left(\frac{l\sin\theta}{k_{B}T}\sqrt{F_{x}^{2} + F_{y}^{2}}\right) 
 \sin\theta \: \df{\theta}.
\end{equation}
We need to study two different situations $K_{1} < K_{2}$ and 
$K_{1} > K_{2}$. Note that the exponential in the integral for the partition 
function above has its peak near the center $\theta = \frac{\pi}{2}$ when 
$K_{1} - K_{2} < 0$, and near $\theta = 0,\pi$ when $K_{1} - K_{2} > 0$. We 
will consider these cases separately since they lead to qualitatively 
different electro-mechanical responses. 

\paragraph{Case 1a:} In this case $K_{1} - K_{2} < 0$ and the exponential 
in the integral in \eqref{eq:blue} is maximum near 
$\theta = \frac{\pi}{2}$. So, we change variable to 
$\alpha = \frac{\pi}{2} - \theta$, then the integral becomes 
\begin{equation}
 Z_{1} = \int_{-\pi/2}^{\pi/2}\exp\left(-\frac{-\frac{K_{1} - K_{2}}{2} E_{z}^{2}
 \sin^{2}\alpha - \frac{K_{2}}{2}E_{z}^{2} - F_{z}l\sin\alpha}{k_{B}T}\right)
 I_{0}\left(\frac{l\cos\alpha}{k_{B}T}\sqrt{F_{x}^{2} + F_{y}^{2}}\right) 
 \cos\alpha \: \df{\alpha}
\end{equation}
For small values of $F_{z}l$ and large values of $E_{z}$ the exponential will
be maximum near $\alpha = 0$ if $K_{1} - K_{2} < 0$. We assume 
$\frac{l\cos\alpha}{k_{B}T}\sqrt{F_{x}^{2} + F_{y}^{2}}$ is small 
and expand 
the modified Bessel function near $0$ as $I_{0}(z)\approx 1 + \frac{z^{2}}{4}$,
then with $y=\sin\alpha$, $dy=\cos\alpha\,d\alpha$, we get 
\begin{equation}
\begin{split}
 Z_{1} = & \exp\left(\frac{K_{2}E_{z}^{2}}{2k_{B}T}\right)
 \left[1 + \frac{l^{2}}{4k_{B}^{2}T^{2}}(F_{x}^{2} + F_{y}^{2})\right]
 \int_{-1}^{1} \exp\left(-\frac{(-K_{1}+K_{2})E_{z}^{2}y^{2} 
 - 2F_{z}ly}{2k_{B}T}\right) \df{y} \\
 = & \exp\left(\frac{K_{2}E_{z}^{2}}{2k_{B}T}\right)\left[1 
 + \frac{l^{2}}{4k_{B}^{2}T^{2}}(F_{x}^{2} + F_{y}^{2})\right]\frac{1}{2E_{z}}
 \sqrt{\frac{2\pi k_{B}T}{-K_{1} + K_{2}}}\exp\left(\frac{F_{z}^{2}l^{2}}
 {2k_{B}T(-K_{1}+K_{2})E_{z}^{2}}\right) \times \\
   &\left[\erf\left(\frac{\sqrt{K_{2}-K_{1}}E_{z}}{\sqrt{2k_{B}T}} 
 - \frac{F_{z}l}{\sqrt{2k_{B}T}\sqrt{K_{2}-K_{1}}E_{z}}\right)
  + \erf\left(\frac{\sqrt{K_{2}-K_{1}}E_{z}}{\sqrt{2k_{B}T}} + \frac{F_{z}l}
 {\sqrt{2k_{B}T}\sqrt{K_{2}-K_{1}}E_{z}}\right)\right].
\end{split}
\end{equation}
This expression for $Z_{1}$ should be valid for any $E_{z} > 0$, small 
$\frac{\sqrt{F_{x}^{2} + F_{y}^{2}}l}{k_{B}T}$, but any value of $F_{z}$. 
The free energy is:
\begin{eqnarray}
 G(\mathbf{F},\mathbf{E},T) & =& -Nk_{B}T\log Z_{1} \nonumber \\
 & =& -\frac{NK_{2}E_{z}^{2}}{2} 
 - Nk_{B}T\log\left[1 + \frac{l^{2}(F_{x}^{2} + F_{y}^{2})}{4k_{B}^{2}T^{2}}\right] 
 + \frac{Nk_{B}T}{2}\log\frac{(-K_{1}+K_{2})E_{z}^{2}}{2\pi k_{B}T}
 - \frac{NF_{z}^{2}l^{2}}{2(-K_{1}+K_{2})E_{z}^{2}} \nonumber \\  
 &  &-Nk_{B}T\log\Bigg[\erf\left(\frac{\sqrt{K_{2}-K_{1}}E_{z}}{\sqrt{2k_{B}T}} 
 - \frac{F_{z}l}{\sqrt{2k_{B}T}\sqrt{K_{2}-K_{1}}E_{z}}\right) \nonumber \\
 &  & \qquad \qquad \qquad + \erf\left(\frac{\sqrt{K_{2}-K_{1}}E_{z}}{\sqrt{2k_{B}T}} + \frac{F_{z}l}{\sqrt{2k_{B}T}\sqrt{K_{2}-K_{1}}E_{z}}\right)\Bigg]. \label{eq:G1a}
\end{eqnarray}  
An important point about the expression for the free energy above is that 
it exhibits a transverse isotropic symmetry with the $z$-axis aligned with 
the electric field vector $\mathbf{E}$ as the symmetry axis. Since 
\begin{equation}
 E_{z} = \sqrt{\mathbf{E}\cdot\mathbf{E}} = |\mathbf{E}|, \qquad
 F_{z} = \frac{\mathbf{F}\cdot\mathbf{E}}{|\mathbf{E}|}, \quad   
 F_{x}^{2} + F_{y}^{2} = |\mathbf{F} 
 - \frac{\mathbf{F}\cdot\mathbf{E}}{|\mathbf{E}|}|^2 
 = \mathbf{F}\cdot\mathbf{F} - \frac{(\mathbf{F}\cdot\mathbf{E})^{2}}
 {\mathbf{E}\cdot\mathbf{E}},
\end{equation}
the free energy depends only on the scalars $\mathbf{E}\cdot\mathbf{E}$,
$\mathbf{E}\cdot\mathbf{F}$ and $\mathbf{F}\cdot\mathbf{F}$, as expected.

\begin{figure}
\centering
\includegraphics{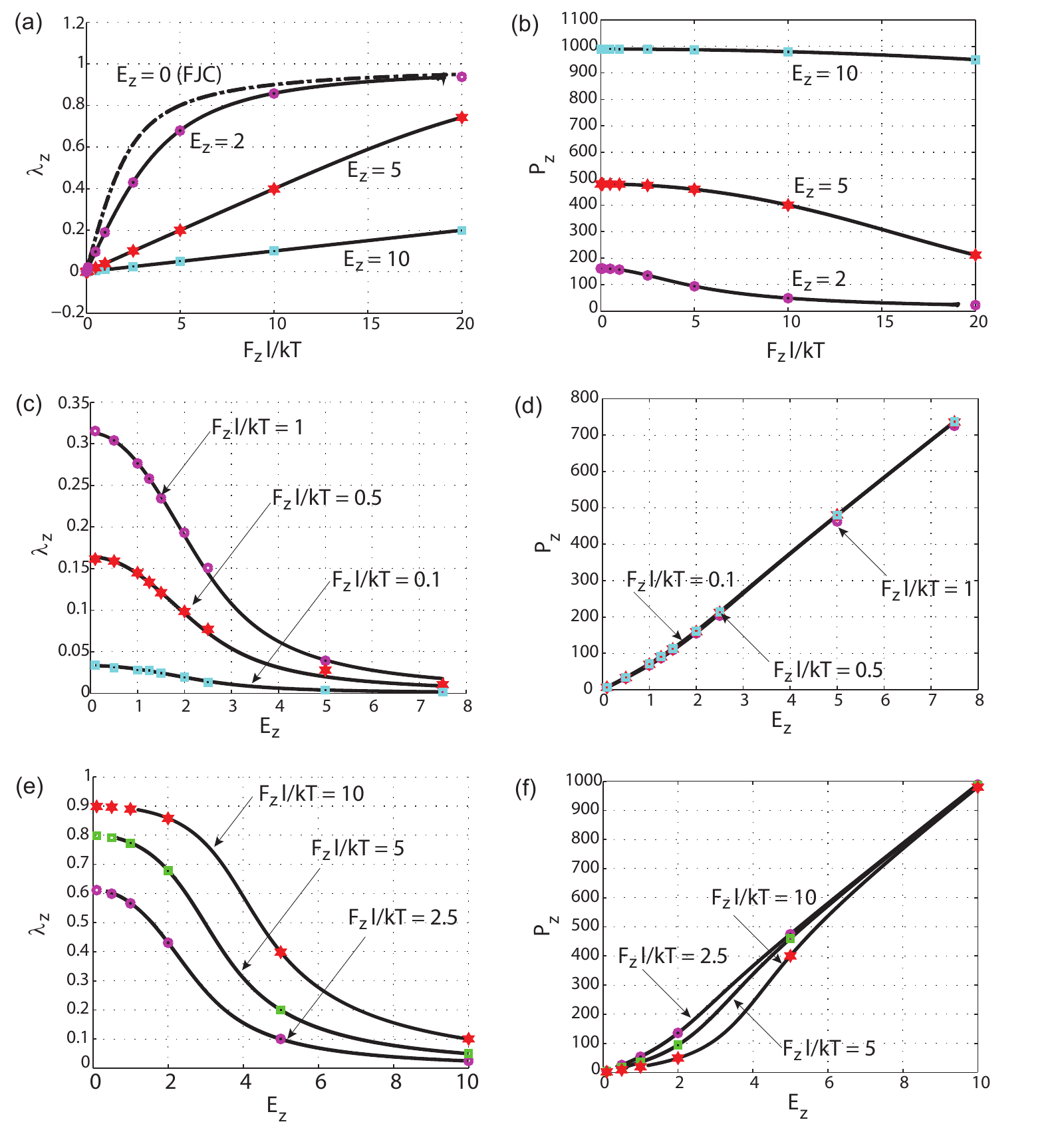}
\caption{Stretch $\lambda_{z}$ and polarization $P_{z}$ for various values of
$F_{z}$ and $E_{z}$ obtained from differentiating the free energy in 
eqn. (\ref{eq:G1a}). We assume $N = 100$, $K_{1}=0$, $K_{2}=1$ and $k_{B}T=1$ for all 
panels. The solid lines are results of analytical formulae eqn. (\ref{eq:lamzG1a}) and 
eqn. (\ref{eq:PzG1a}). In some panels (e.g., panel (e)) the solid lines do 
not continue to the end because of numerical difficulties with the $\erf$ function. 
In panels (a) and (b) the symbols are results from Monte 
Carlo simulations -- $E_{z} = 2$ magenta circles, $E_{z} = 5$ red stars, 
$E_{z} = 10$ cyan squares. The dashed line shows the force-stretch relation for a
freely jointed chain under zero electric field. In panels (c) and (d) the symbols are 
results from Monte Carlo simulations -- $\frac{F_{z}l}{k_{B}T} = 0.1$ cyan squares,
$\frac{F_{z}l}{k_{B}T} = 0.5$ red stars, $\frac{F_{z}l}{k_{B}T} = 1$ magenta
circles. In panels (e) and (f) the symbols are results of Monte Carlo simulations
-- $\frac{F_{z}l}{k_{B}T} = 2.5$ magenta circles, $\frac{F_{z}l}{k_{B}T} = 5$, green 
squares and $\frac{F_{z}l}{k_{B}T} = 10$ red stars.}
\label{fig:fig1K1lK2}
\end{figure}

By differentiating the expression for the free energy we get for $\lambda_{z}$
\begin{equation} \label{eq:lamzG1a} 
\begin{split}
 \lambda_{z} =& \frac{F_{z}l}{(K_{2}-K_{1})E_{z}^{2}} 
 + \sqrt{\frac{2k_{B}T}{\pi(K_{2}-K_{1})E_{z}^{2}}}\times \\
 & \left[\frac{-\exp\left(-(\frac{\sqrt{K_{2}-K_{1}}E_{z}}{\sqrt{2k_{B}T}} 
 - \frac{F_{z}l}{\sqrt{2k_{B}T}\sqrt{K_{2}-K_{1}}E_{z}})^{2}\right)
  + \exp\left(-(\frac{\sqrt{K_{2}-K_{1}}E_{z}}{\sqrt{2k_{B}T}} 
 + \frac{F_{z}l}{\sqrt{2k_{B}T}\sqrt{K_{2}-K_{1}}E_{z}})^{2}\right)} 
 {\erf\left(\frac{\sqrt{K_{2}-K_{1}}E_{z}}{\sqrt{2k_{B}T}} 
 - \frac{F_{z}l}{\sqrt{2k_{B}T}\sqrt{K_{2}-K_{1}}E_{z}}\right) 
 + \erf\left(\frac{\sqrt{K_{2}-K_{1}}E_{z}}{\sqrt{2k_{B}T}} + \frac{F_{z}l}
 {\sqrt{2k_{B}T}\sqrt{K_{2}-K_{1}}E_{z}}\right)}\right]. 
\end{split}
\end{equation}
And, for the polarization $P_{z}$, we get
\begin{equation} \label{eq:PzG1a}
\begin{split}
 &P_{z} = NK_{2}E_{z} - \frac{Nk_{B}T}{E_{z}} 
 - \frac{NF_{z}^{2}l^{2}}{(K_{2}-K_{1})E_{z}^{3}} \\ 
 &+ \frac{N\sqrt{2k_{B}T(K_{2}-K_{1})}}{\sqrt{\pi}} 
 \left[\frac{\exp\left(-(\frac{\sqrt{K_{2}-K_{1}}E_{z}}{\sqrt{2k_{B}T}} 
 - \frac{F_{z}l}{\sqrt{2k_{B}T}\sqrt{K_{2}-K_{1}}E_{z}})^{2}\right)
  + \exp\left(-(\frac{\sqrt{K_{2}-K_{1}}E_{z}}{\sqrt{2k_{B}T}} 
 + \frac{F_{z}l}{\sqrt{2k_{B}T}\sqrt{K_{2}-K_{1}}E_{z}})^{2}\right)} 
 {\erf\left(\frac{\sqrt{K_{2}-K_{1}}E_{z}}{\sqrt{2k_{B}T}} 
 - \frac{F_{z}l}{\sqrt{2k_{B}T}\sqrt{K_{2}-K_{1}}E_{z}}\right) 
 + \erf\left(\frac{\sqrt{K_{2}-K_{1}}E_{z}}{\sqrt{2k_{B}T}} + \frac{F_{z}l}
 {\sqrt{2k_{B}T}\sqrt{K_{2}-K_{1}}E_{z}}\right)}\right] \\
 &+ \frac{N\sqrt{2k_{B}T}F_{z}l}{\sqrt{\pi}\sqrt{K_{2}-K_{1}}E_{z}^{2}}
 \left[\frac{\exp\left(-(\frac{\sqrt{K_{2}-K_{1}}E_{z}}{\sqrt{2k_{B}T}} 
 - \frac{F_{z}l}{\sqrt{2k_{B}T}\sqrt{K_{2}-K_{1}}E_{z}})^{2}\right)
  - \exp\left(-(\frac{\sqrt{K_{2}-K_{1}}E_{z}}{\sqrt{2k_{B}T}} 
 + \frac{F_{z}l}{\sqrt{2k_{B}T}\sqrt{K_{2}-K_{1}}E_{z}})^{2}\right)} 
 {\erf\left(\frac{\sqrt{K_{2}-K_{1}}E_{z}}{\sqrt{2k_{B}T}} 
 - \frac{F_{z}l}{\sqrt{2k_{B}T}\sqrt{K_{2}-K_{1}}E_{z}}\right) 
 + \erf\left(\frac{\sqrt{K_{2}-K_{1}}E_{z}}{\sqrt{2k_{B}T}} + \frac{F_{z}l}
 {\sqrt{2k_{B}T}\sqrt{K_{2}-K_{1}}E_{z}}\right)}\right]
\end{split}
\end{equation}
The remaining stretches and polarizations are:
\begin{equation} \label{eq:lamxK1lK2}
 \lambda_{x} = \frac{r_{x}}{Nl} = \frac{F_{x}l}{2k_{B}T\left(1
 + \frac{l^{2}\left((F_{x}^{2}+F_{y}^{2}\right)}{4k_{B}^{2}T^{2}}\right)}, \qquad 
 \lambda_{y} = \frac{r_{y}}{Nl} = \frac{F_{y}l}{2k_{B}T\left(1
 + \frac{l^{2}\left(F_{x}^{2}+F_{y}^{2}\right)}{4k_{B}^{2}T^{2}}\right)}, \qquad 
 P_{x} = 0, \qquad 
 P_{y} = 0.
\end{equation}
In practice the two $\erf$ terms in the denominators of the expressions
above can sum to near zero for large values of $F_{z}$ or small values of 
$E_{z}$ so we encounter some numerical difficulties in the plots. 
In figure~\ref{fig:fig1K1lK2} we show that the above expressions for 
$\lambda_{z}$ and $P_{z}$ match those obtained from Monte Carlo simulations 
quite well for a large range of $F_{z}$ and $E_{z}$. From 
figure~\ref{fig:fig1K1lK2}(a) it is apparent that when $K_{1} < K_{2}$ and
$\mathbf{E}_{z} \neq 0$ the stretches $\lambda_{z}$ are all less than 
those predicted by the FJC formula (which assumes $\mathbf{E} = 0$). As
$E_{z}$ increases at a fixed $F_{z}$ the $\lambda_{z}$ progressively
decreases. It is also easy to verify from figure~\ref{fig:fig1K1lK2}(c)
and (e) that as $E_{z} \rightarrow 0$ we recover the FJC values of 
$\lambda_{z}$ for all $F_{z}$.  

\begin{figure}
\centering
\includegraphics{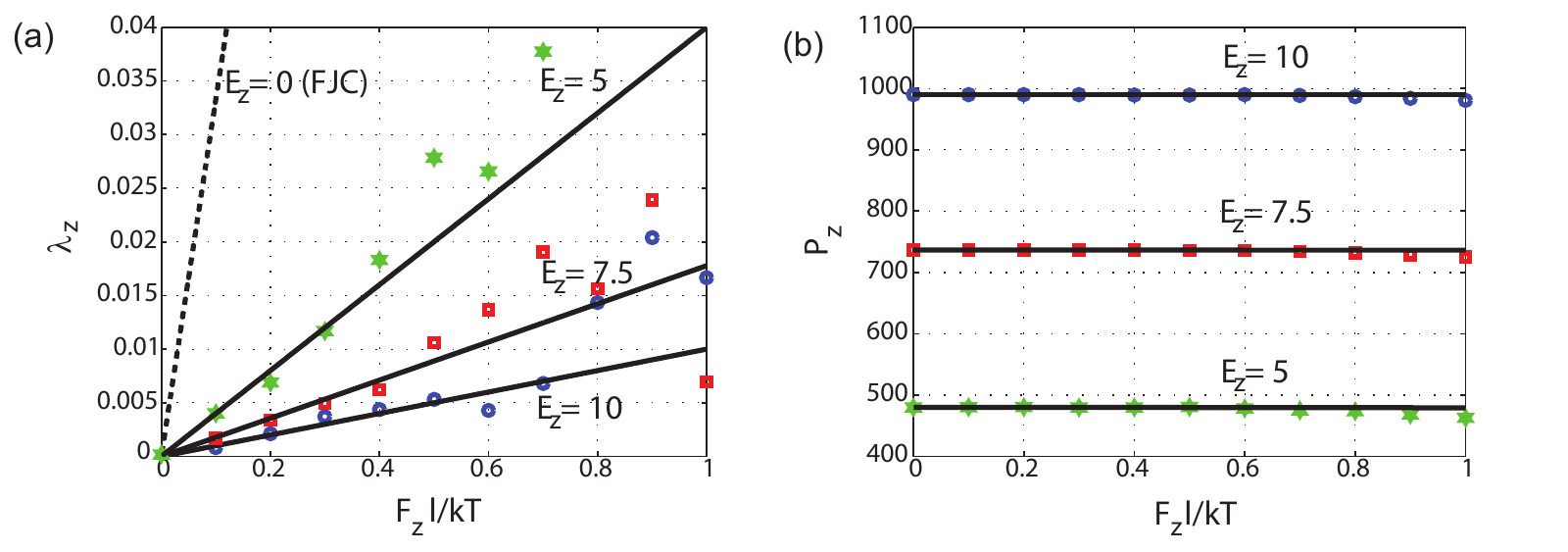}
\caption{Stretch $\lambda_{z}$ and polarization $P_{z}$ for various values of
$F_{z}$ and $E_{z}$ obtained from differentiating the free energy in 
eqn. (\ref{eq:gibbs1}). We assume $N=100$, $K_{1}=0$, $K_{2}=1$ and $k_{B}T=1$ for all 
panels. The lines are results of analytical formulae eqn. (\ref{eq:lamzgibbs1})
and eqn. (\ref{eq:Pzgibbs1}). In panels (a) and (b) the symbols are results 
from Monte Carlo simulations -- $E_{z} = 5$ green stars.  $E_{z} = 7.5$ red
squares, $E_{z} = 10$ blue circles. The dashed line in panel (a) is the
prediction of the FJC formula which assumes zero electric field.} 
\label{fig:fig4gauK1lK2}
\end{figure}

It is useful to consider the limit of large $E_{z}$ in which 
$(K_{2}-K_{1})E_{z}^{2} \gg F_{z}l$ so that the $\erf$ functions in 
eqn. (\ref{eq:G1a}) go to 1. This is equivalent to changing the limits
to $\pm\infty$ in the integral for the partition function so that  
\begin{equation}
\begin{split}
 Z_{1} \approx & \exp\left(\frac{K_{2}E_{z}^{2}}{2k_{B}T}\right)
 \left[1 + \frac{l^{2}}{4k_{B}^{2}T^{2}}(F_{x}^{2} + F_{y}^{2})\right]
 \int_{-\infty}^{\infty}
 \exp\left(-\frac{(-K_{1}+K_{2})E_{z}^{2}y^{2} 
 - 2F_{z}ly}{2k_{B}T}\right) \df{y} \\
 =& \exp\left(\frac{K_{2}E_{z}^{2}}{2k_{B}T}\right)\left[1 
 + \frac{l^{2}}{4k_{B}^{2}T^{2}}(F_{x}^{2} + F_{y}^{2})\right]\frac{1}{E_{z}}
 \sqrt{\frac{2\pi k_{B}T}{-K_{1} + K_{2}}}\exp\left(\frac{F_{z}^{2}l^{2}}
 {2k_{B}T(-K_{1}+K_{2})E_{z}^{2}}\right)
\end{split}
\end{equation}
Then, the free energy of a single polymer chain is
\begin{equation}
 G(\mathbf{F},\mathbf{E},T) = -\frac{NK_{2}E_{z}^{2}}{2} 
 - Nk_{B}T\log\left[1 + \frac{l^{2}(F_{x}^{2} + F_{y}^{2})}{4k_{B}^{2}T^{2}}\right] 
 + \frac{Nk_{B}T}{2}\log\frac{(-K_{1}+K_{2})E_{z}^{2}}{2\pi k_{B}T}
 - \frac{NF_{z}^{2}l^{2}}{2(-K_{1}+K_{2})E_{z}^{2}}. \label{eq:gibbs1}
\end{equation}  
The expressions for the stretches and polarizations simplify considerably in
this limit:
\begin{align} \label{eq:Pzgibbs1}
 \lambda_{x} &= \frac{F_{x}l}{2k_{B}T\left(1
 + \frac{l^{2}(F_{x}^{2}+F_{y}^{2})}{4k_{B}^{2}T^{2}}\right)}, \qquad 
 \lambda_{y} = \frac{F_{y}l}{2k_{B}T\left(1
 + \frac{l^{2}(F_{x}^{2}+F_{y}^{2})}{4k_{B}^{2}T^{2}}\right)}, \qquad 
 \lambda_{z} = \frac{F_{z}l}{(K_{2} - K_{1})E_{z}^{2}}, \\
\label{eq:lamzgibbs1}
P_{x} &= 0, \qquad P_{y} = 0, \qquad 
 P_{z} = NK_{2}E_{z} - \frac{Nk_{B}T}{E_{z}} 
 - \frac{NF_{z}^{2}l^{2}}{(K_{2}-K_{1})E_{z}^{3}}.
\end{align} 
These formulae are useful in the limit of large $E_{z}$ and small $F_{z}$ 
in which we have numerical difficulties with the $\erf$ function. Some such 
results are plotted in figure~\ref{fig:fig4gauK1lK2} for $E_{z} = 5,7.5,10$
for $\frac{F_{z}l}{k_{B}T} \leq 1$. It is clear that the trends in 
$\lambda_{z}$ and $P_{z}$ are correctly captured at larger $E_{z}$.

\begin{figure}
\centering
\includegraphics{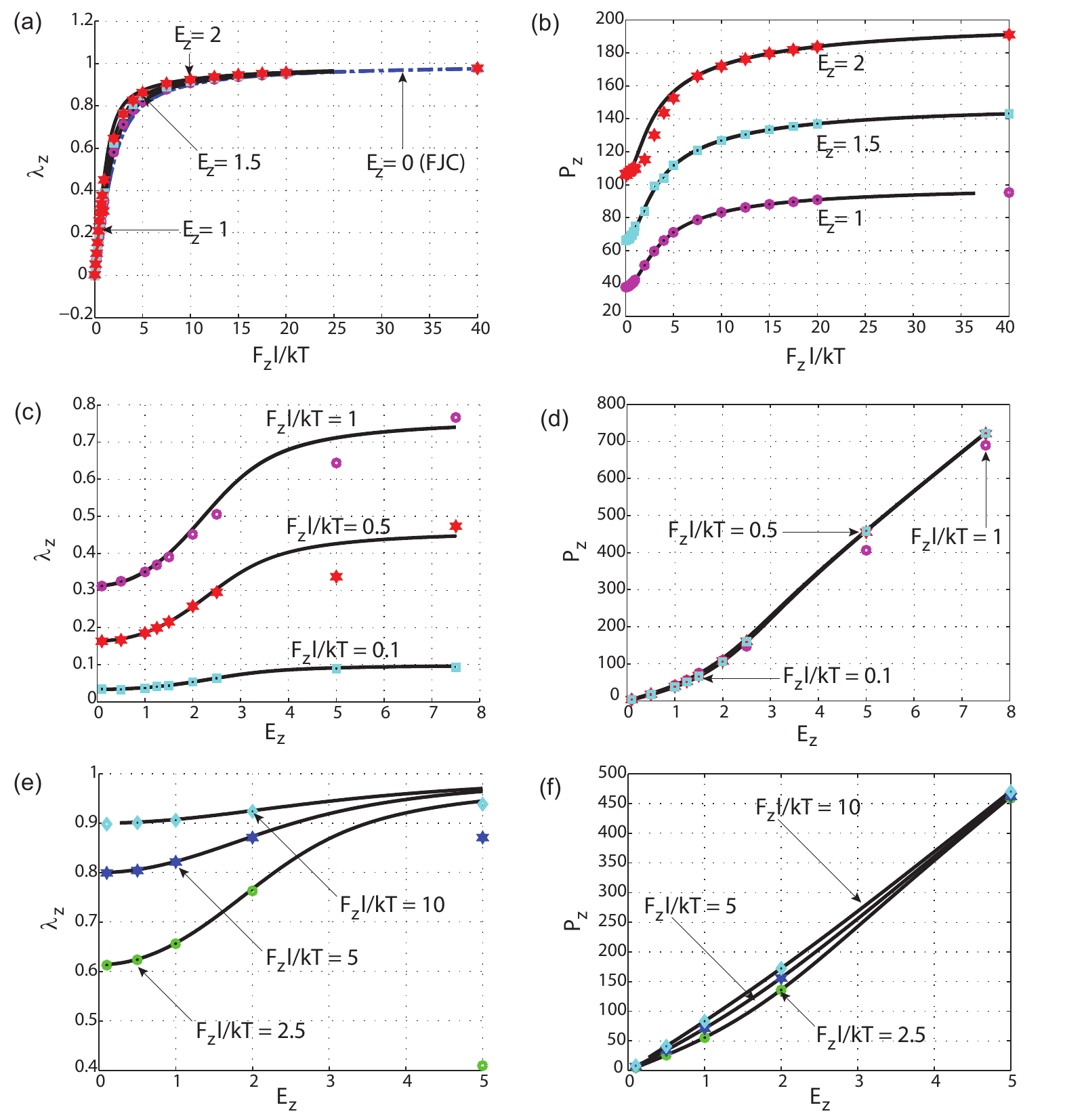}
\caption{Stretch $\lambda_{z}$ and polarization $P_{z}$ for various values of
$F_{z}$ and $E_{z}$ obtained from differentiating the free energy in 
eqn. (\ref{eq:G1b}). We assume $N=100$, $K_{1}=1$, $K_{2}=0$ and $k_{B}T=1$ for all 
panels. The lines are results of analytical formulae eqn.(\ref{eq:lamzG1b}) and 
eqn. (\ref{eq:PzG1b}). In panels (a) and (b) the symbols are results from Monte 
Carlo simulations -- $E_{z} = 1$ magenta circles, $E_{z} = 1.5$ cyan squares, 
$E_{z} = 2$ red stars. The blue dashed line in panel (a) is the force-stretch
relation of the freely jointed chain under zero electric field. 
In panels (c) and (d) the symbols are 
results from Monte Carlo simulations -- $\frac{F_{z}l}{k_{B}T} = 0.1$ cyan squares,
$\frac{F_{z}l}{k_{B}T} = 0.5$ red stars, $\frac{F_{z}l}{k_{B}T} = 1$ magenta
circles. In panels (e) and (f) the symbols are 
results from Monte Carlo simulations -- $\frac{F_{z}l}{k_{B}T} = 2.5$ green circles, 
$\frac{F_{z}l}{k_{B}T} = 5$ blue stars, $\frac{F_{z}l}{k_{B}T} = 10$ cyan diamonds.} 
\label{fig:fig2K2lK1}
\end{figure}

\paragraph{Case 1b:} Now let us consider the case when $K_{1} - K_{2} > 0$, 
so that the exponential in the integral in eqn. (\ref{eq:blue}) has its maxima 
near the boundaries $\theta = 0,\pi$. Assuming
$\frac{l\sin\theta\sqrt{F_{x}^{2} + F_{y}^{2}}}{k_{B}T} < 1$, we again expand
the modified Bessel function $I_{0}(z) \approx 1 + \frac{z^{2}}{4}$. Then,
\begin{equation}
 Z_{1} = \int_{0}^{\pi}\exp\left(-\frac{\frac{-K_{1} + K_{2}}{2} E_{z}^{2}
 \cos^{2}\theta - \frac{K_{2}}{2}E_{z}^{2} - F_{z}l\cos\theta}{k_{B}T}\right)
 \left[1 + \frac{(F_{x}^{2}+F_{y}^{2})l^{2}}{4k_{B}^{2}T^{2}}\sin^{2}\theta
 \right] \sin\theta \: \df{\theta}. 
\end{equation}
In fact, this approximation for the Bessel function is valid as long as
$l\sqrt{F_{x}^{2} + F_{y}^{2}} < k_{B}T$, irrespective of where the maxima 
of the exponential are (i.e., irrespective of the sign of $K_{1}-K_{2}$). 
Now take $y = -\cos\theta$, then $dy = \sin\theta d\theta$ and the integral
changes to
\begin{equation}
 Z_{1} = \int_{-1}^{1}\exp\left(-\frac{\frac{-K_{1} + K_{2}}{2} E_{z}^{2}y^{2}
 - \frac{K_{2}}{2}E_{z}^{2} + F_{z}ly}{k_{B}T}\right)
 \left[1 + \frac{(F_{x}^{2}+F_{y}^{2})l^{2}}{4k_{B}^{2}T^{2}}(1-y^{2})\right] \: \df{y}
\end{equation}
Neglecting terms higher than linear order in $y$ in the square brackets, we
write this as 
\begin{equation} \label{eq:Z1corre} 
 Z_{1} = \left[1 + \frac{(F_{x}^{2}+F_{y}^{2})l^{2}}{4k_{B}^{2}T^{2}}\right]
 \exp\left(\frac{K_{2}E_{z}^{2}}{2k_{B}T}\right) 
 \int_{-1}^{1}\exp\left(-\frac{\frac{-K_{1} + K_{2}}{2} E_{z}^{2}y^{2} 
 + F_{z}ly}{k_{B}T}\right) \: \df{y}
\end{equation}
Integrating out $y$ we get
\begin{equation} \label{eq:diffclt}
\begin{split}
 Z_{1} =& \left[1 + \frac{(F_{x}^{2}+F_{y}^{2})l^{2}}{4k_{B}^{2}T^{2}}\right]
 \exp\left(\frac{K_{2}E_{z}^{2}}{2k_{B}T}\right)\frac{1}{2E_{z}}
 \sqrt{\frac{2\pi k_{B}T}{-K_{2}+K_{1}}}
 \exp\left(-\frac{F_{z}^{2}l^{2}}{2k_{B}TE_{z}^{2}(K_{1}-K_{2})}\right) \times \\
   &\left[\erfi\left(\frac{1}{E_{z}}\sqrt{\frac{k_{B}T}{2(K_{1}-K_{2})}}
  \left[\frac{(K_{1}-K_{2})E_{z}^{2} - F_{z}l}{k_{B}T}\right]\right)  
 + \erfi\left(\frac{1}{E_{z}}\sqrt{\frac{k_{B}T}{2(K_{1}-K_{2})}}\left[
 \frac{(K_{1}-K_{2})E_{z}^{2} + F_{z}l}{k_{B}T}\right]\right)\right].  
\end{split}
\end{equation}
The free energy is, then
\begin{eqnarray}
 G(\mathbf{F},\mathbf{E},T) & =& -\frac{NK_{2}E_{z}^{2}}{2} 
 - Nk_{B}T\log\left[1 + \frac{l^{2}(F_{x}^{2} + F_{y}^{2})}{4k_{B}^{2}T^{2}}\right] 
 + \frac{Nk_{B}T}{2}\log\frac{(K_{1}-K_{2})E_{z}^{2}}{2\pi k_{B}T}
 + \frac{NF_{z}^{2}l^{2}}{2(K_{1}-K_{2})E_{z}^{2}} \nonumber \\  
 &  &-Nk_{B}T\log\Bigg[\erfi\left(\frac{\sqrt{K_{1}-K_{2}}E_{z}}{\sqrt{2k_{B}T}} 
 - \frac{F_{z}l}{\sqrt{2k_{B}T}\sqrt{K_{1}-K_{2}}E_{z}}\right) \nonumber \\
 &  & \qquad \qquad \qquad + \erfi\left(\frac{\sqrt{K_{1}-K_{2}}E_{z}}{\sqrt{2k_{B}T}} 
 + \frac{F_{z}l}{\sqrt{2k_{B}T}\sqrt{K_{1}-K_{2}}E_{z}}\right)\Bigg].
 \label{eq:G1b}
\end{eqnarray}  
By differentiation, the stretch $\lambda_{z}$ is
\begin{equation}  \label{eq:lamzG1b}
\begin{split}
 \lambda_{z} =& -\frac{F_{z}l}{(K_{1}-K_{2})E_{z}^{2}} 
 + \sqrt{\frac{2k_{B}T}{\pi(K_{1}-K_{2})E_{z}^{2}}}\times \\
 & \left[\frac{-\exp\left((\frac{\sqrt{K_{1}-K_{2}}E_{z}}{\sqrt{2k_{B}T}} 
 - \frac{F_{z}l}{\sqrt{2k_{B}T}\sqrt{K_{1}-K_{2}}E_{z}})^{2}\right)
  + \exp\left((\frac{\sqrt{K_{1}-K_{2}}E_{z}}{\sqrt{2k_{B}T}} 
 + \frac{F_{z}l}{\sqrt{2k_{B}T}\sqrt{K_{1}-K_{2}}E_{z}})^{2}\right)} 
 {\erfi\left(\frac{\sqrt{K_{1}-K_{2}}E_{z}}{\sqrt{2k_{B}T}} 
 - \frac{F_{z}l}{\sqrt{2k_{B}T}\sqrt{K_{1}-K_{2}}E_{z}}\right) 
 + \erfi\left(\frac{\sqrt{K_{1}-K_{2}}E_{z}}{\sqrt{2k_{B}T}} + \frac{F_{z}l}
 {\sqrt{2k_{B}T}\sqrt{K_{1}-K_{2}}E_{z}}\right)}\right].
\end{split}
\end{equation}
The polarization $P_{z}$ is
\begin{equation}
\begin{split}  \label{eq:PzG1b}
 &P_{z} = NK_{2}E_{z} - \frac{Nk_{B}T}{E_{z}} 
 + \frac{NF_{z}^{2}l^{2}}{(K_{1}-K_{2})E_{z}^{3}} \\ 
 &+ \frac{N\sqrt{2k_{B}T(K_{1}-K_{2})}}{\sqrt{\pi}} 
 \left[\frac{\exp\left((\frac{\sqrt{K_{1}-K_{2}}E_{z}}{\sqrt{2k_{B}T}} 
 - \frac{F_{z}l}{\sqrt{2k_{B}T}\sqrt{K_{1}-K_{2}}E_{z}})^{2}\right)
  + \exp\left((\frac{\sqrt{K_{1}-K_{2}}E_{z}}{\sqrt{2k_{B}T}} 
 + \frac{F_{z}l}{\sqrt{2k_{B}T}\sqrt{K_{1}-K_{2}}E_{z}})^{2}\right)} 
 {\erfi\left(\frac{\sqrt{K_{1}-K_{2}}E_{z}}{\sqrt{2k_{B}T}} 
 - \frac{F_{z}l}{\sqrt{2k_{B}T}\sqrt{K_{1}-K_{2}}E_{z}}\right) 
 + \erfi\left(\frac{\sqrt{K_{1}-K_{2}}E_{z}}{\sqrt{2k_{B}T}} + \frac{F_{z}l}
 {\sqrt{2k_{B}T}\sqrt{K_{1}-K_{2}}E_{z}}\right)}\right] \\
 &+ \frac{N\sqrt{2k_{B}T}F_{z}l}{\sqrt{\pi}\sqrt{K_{1}-K_{2}}E_{z}^{2}}
 \left[\frac{\exp\left((\frac{\sqrt{K_{1}-K_{2}}E_{z}}{\sqrt{2k_{B}T}} 
 - \frac{F_{z}l}{\sqrt{2k_{B}T}\sqrt{K_{1}-K_{2}}E_{z}})^{2}\right)
  - \exp\left((\frac{\sqrt{K_{1}-K_{2}}E_{z}}{\sqrt{2k_{B}T}} 
 + \frac{F_{z}l}{\sqrt{2k_{B}T}\sqrt{K_{1}-K_{2}}E_{z}})^{2}\right)} 
 {\erfi\left(\frac{\sqrt{K_{1}-K_{2}}E_{z}}{\sqrt{2k_{B}T}} 
 - \frac{F_{z}l}{\sqrt{2k_{B}T}\sqrt{K_{1}-K_{2}}E_{z}}\right) 
 + \erfi\left(\frac{\sqrt{K_{1}-K_{2}}E_{z}}{\sqrt{2k_{B}T}} + \frac{F_{z}l}
 {\sqrt{2k_{B}T}\sqrt{K_{1}-K_{2}}E_{z}}\right)}\right].
\end{split} 
\end{equation}
The remaining stretches and polarizations are:
\begin{equation} \label{eq:lamxK2lK1}
 \lambda_{x} = \frac{r_{x}}{Nl} = \frac{F_{x}l}{2k_{B}T\left(1
 + \frac{l^{2}(F_{x}^{2}+F_{y}^{2})}{4k_{B}^{2}T^{2}}\right)}, \qquad 
 \lambda_{y} = \frac{r_{y}}{Nl} = \frac{F_{y}l}{2k_{B}T\left(1
 + \frac{l^{2}(F_{x}^{2}+F_{y}^{2})}{4k_{B}^{2}T^{2}}\right)}, \qquad 
 P_{x} = 0, \qquad 
 P_{y} = 0.
\end{equation}
These expressions agree quite well with our Monte Carlo results for 
$\lambda_{z}$ and $P_{z}$ over a broad range of $E_{z}$ for 
$\frac{F_{z}l}{k_{B}T} \leq 2$ as shown in figure~\ref{fig:fig2K2lK1}. 
From figure~\ref{fig:fig2K2lK1}(a) it can be seen that when $K_{2} < K_{1}$ and
${E}_{z} \neq 0$ the stretches $\lambda_{z}$ are all more than 
those predicted by the FJC formula (which assumes $\mathbf{E} = 0$). This is
the opposite of what we saw in figure~\ref{fig:fig1K1lK2}(a) which 
assumed $K_{1} < K_{2}$. The analytical formulae match the Monte Carlo data 
for $E_{z} \leq 2$ over a broad range of $F_{z}$. For higher 
$E_{z}=5,10$ the formulae match our Monte Carlo results for $\lambda_{z}$ 
and $P_{z}$ for $\frac{F_{z}l}{k_{B}T} \leq 2$ and $\frac{F_{z}l}{k_{B}T} > 15$, 
but not for intermediate $F_{z}$.  

We will now use a different approach to compute the partition function from
eqn.(\ref{eq:Z1corre}) when the electric field $E_{z}$ is strong. 
Note that the energy minima are near $\theta = 0$ and $\theta = \pi$, so the 
link points in the direction of the electric field (positive $z$ direction) or 
opposite to it. This reminds us of the one-dimensional model of the FJC in 
which each link could point to the left or to the right, but no other 
direction in between. In this case the partition function 
eqn.(\ref{eq:Z1corre}) reduces to the following discrete sum (rather than 
an integral) approximately:
\begin{equation}
 Z_{1} = \left[1 + \frac{(F_{x}^{2}+F_{y}^{2})l^{2}}{4k_{B}^{2}T^{2}}\right]
 \left(\exp\left(\frac{\frac{K_{1}}{2}E_{z}^{2} + F_{z}l}{k_{B}T}\right)
 + \exp\left(\frac{\frac{K_{1}}{2}E_{z}^{2} - F_{z}l}{k_{B}T}\right)\right).
\end{equation}
Hence, the approximate Gibbs free energy is:
\begin{equation} \label{eq:tanhfjc} 
 G(\mathbf{F},\mathbf{E},T) = 
 -N\frac{K_{1}}{2}E_{z}^{2} 
 -Nk_{B}T\log\left[1 + \frac{(F_{x}^{2}+F_{y}^{2})l^{2}}{4k_{B}^{2}T^{2}}\right]
 - Nk_{B}T\log \left(2\cosh\left(\frac{F_{z}l}{k_{B}T}\right)\right).
\end{equation}
By differentiation we can get the following simple expressions for the 
stretches and polarizations:
\begin{align} \label{eq:lamztanh}
 \lambda_{x} &= \frac{F_{x}l}{2k_{B}T\left(1
 + \frac{l^{2}(F_{x}^{2}+F_{y}^{2})}{4k_{B}^{2}T^{2}}\right)}, \qquad 
 \lambda_{y} = \frac{F_{y}l}{2k_{B}T\left(1
 + \frac{l^{2}(F_{x}^{2}+F_{y}^{2})}{4k_{B}^{2}T^{2}}\right)}, \qquad 
 \lambda_{z} = \tanh\left(\frac{F_{z}l}{k_{B}T}\right), \\ \label{eq:Pztanh}
 P_{x} &= 0, \qquad 
 P_{y} = 0, \qquad 
 P_{z} = NK_{1}E_{z}.   
\end{align}
\begin{figure}
\centering
\includegraphics{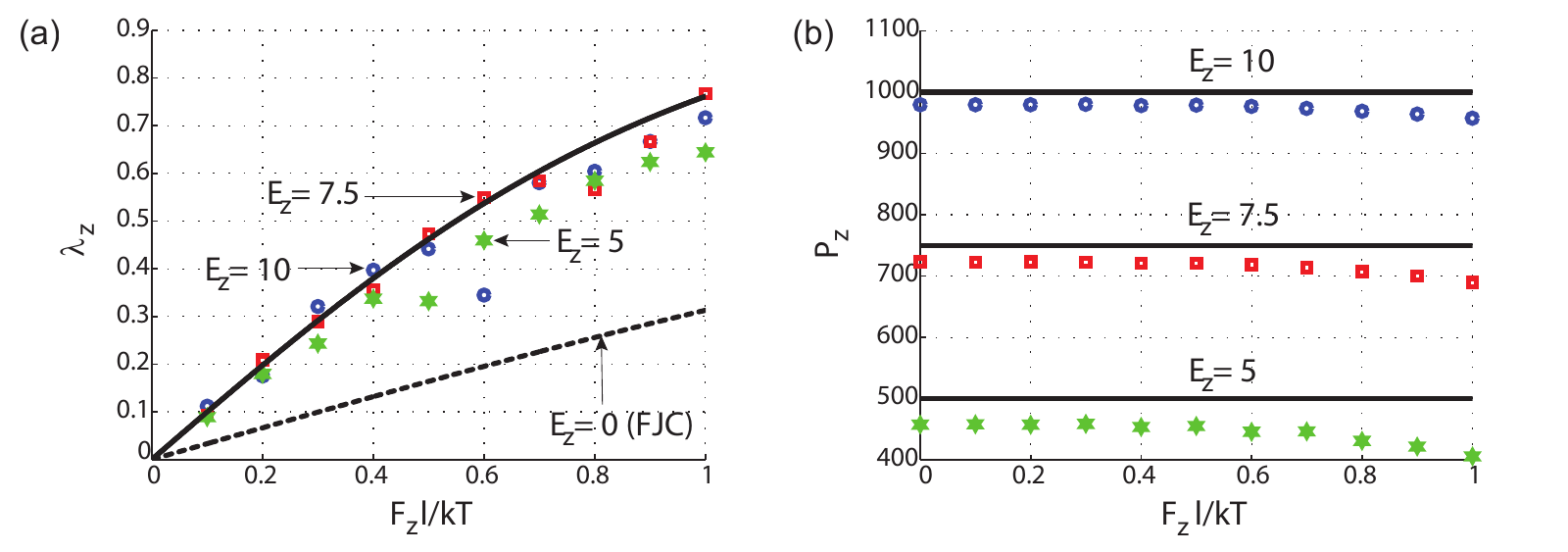}
\caption{Stretch $\lambda_{z}$ and polarization $P_{z}$ for various values of
$F_{z}$ and $E_{z}$ obtained from differentiating the free energy in 
eqn. (\ref{eq:tanhfjc}). We assume $N=100$, $K_{1}=1$, $K_{2}=0$ and $k_{B}T=1$ for all 
panels. The lines are results of analytical formulae eqn. (\ref{eq:lamztanh})
and eqn. (\ref{eq:Pztanh}). In panels (a) and (b) the symbols are results 
from Monte Carlo simulations -- $E_{z} = 5$ green stars.  $E_{z} = 7.5$ red
squares, $E_{z} = 10$ blue circles. The dashed line in panel (a) is the prediction
of the FJC formula which assumes zero electric field.} 
\label{fig:fig5tanhK2lK1}
\end{figure}
The results of these formulae appear in figure~\ref{fig:fig5tanhK2lK1} for
$E_{z} = 5,7.5,10$ and $\frac{F_{z}l}{k_{B}T} \leq 1$. It is apparent that 
for large values of $E_{z}$ the curve
for $\lambda_{z}$ vs. $\frac{F_{z}l}{k_{B}T}$ is independent of $E_{z}$ in 
eqn. (\ref{eq:lamztanh}) as well as the Monte Carlo simulations. 
Eqn. (\ref{eq:Pztanh}) overpredicts the $P_{z}$ compared to the Monte Carlo
simulations, but the estimate gets better as $E_{z}$ increases.

\begin{figure}
\centering
\includegraphics{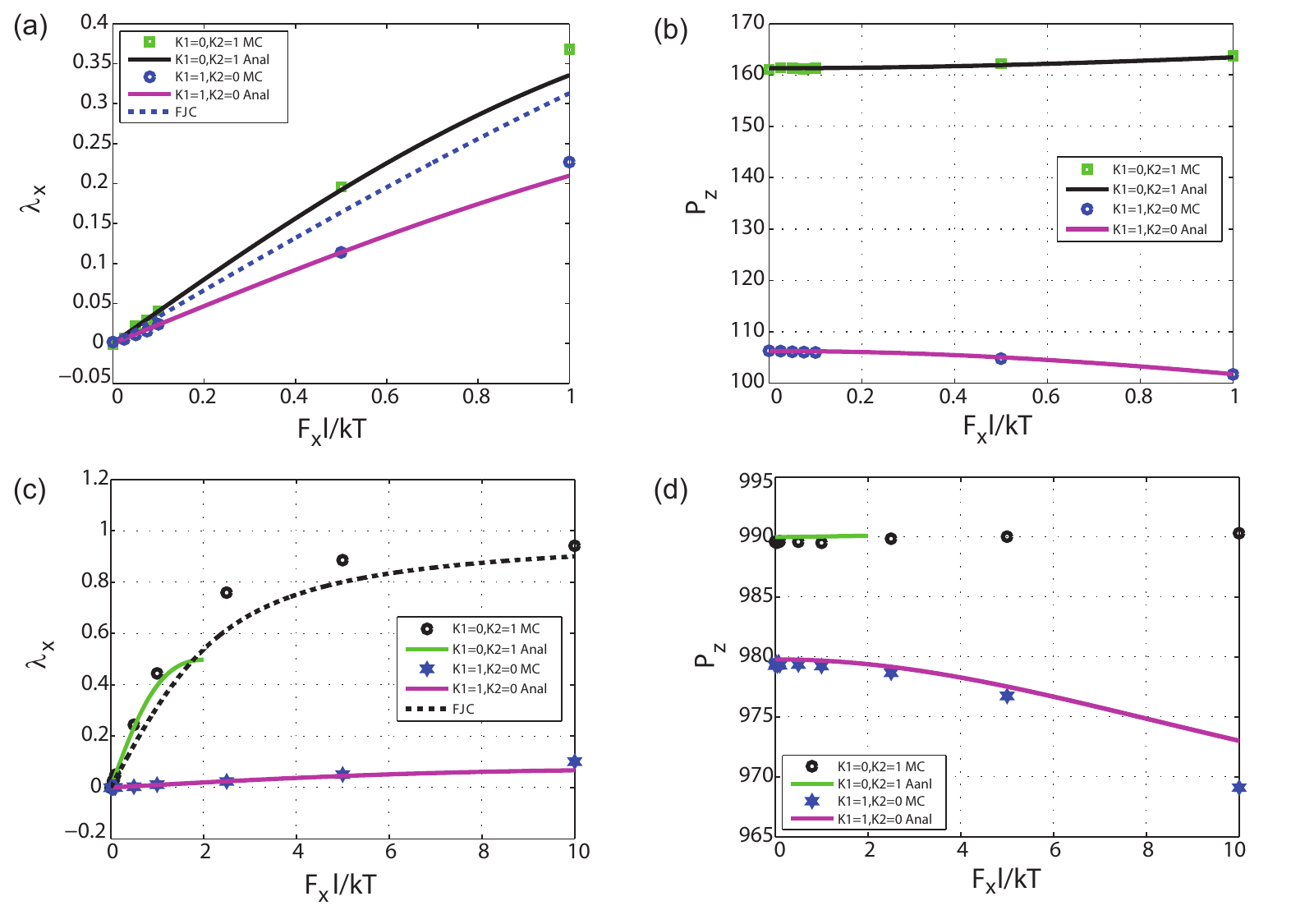}
\caption{Stretch $\lambda_{x}$ and polarization $P_{z}$ for various values of
$F_{x}$ and $E_{z}$ obtained from differentiating the free energies in 
eqn. (\ref{eq:G1aFx}) and eqn. (\ref{eq:G1bFx}). We take $N=100$, $k_{B}T = 1$ and 
consider both cases, $K_{1} = 0, K_{2} = 1$ and $K_{1} = 1, K_{2} = 0$.  
The lines are results of analytical formulae eqn.(\ref{eq:lamxG1aFx}), 
eqn.(\ref{eq:PzG1aFx}), eqn. (\ref{eq:lamxG1bFx}) and eqn. (\ref{eq:PzG1bFx}). 
In panels (a) and (b) the symbols are results from Monte 
Carlo simulations with $E_{z} = 2$ and in panels (c) and (d) with $E_{z} = 10$.
The dashed lines in panels (a) and (c) are the predictions of the FJC formula
which assumes zero electric field.}
\label{fig:fig3EzFx}
\end{figure}

\paragraph{Case 1c:} In many applications of dielectric polymers the electric
field is perpendicular to the direction of stretching of the chains. Yet, in 
the previous two cases we focused primarily on the situation when the electric
field and force are aligned with each other. We could account for the situation
in which the electric field and force are perpendicular to each other by 
setting $E_{z} \neq 0$, $F_{z} = 0$ and $F_{x} \neq 0$ (and/or $F_{y} \neq 0$)
then eqn. (\ref{eq:lamxK1lK2}) and eqn. (\ref{eq:lamxK2lK1}) showed that the 
stretches $\lambda_{x}$ and $\lambda_{y}$ are independent of the electric field.   
However, our Monte Carlo simulations have shown that when $F_{z} = 0$ and 
$F_{x} \neq 0$ the curve of $\lambda_{x}$ vs. $F_{x}$ is affected by $E_{z}$.
Our analysis in cases 1a and 1b does not capture this effect of $E_{z}$ on
$\lambda_{x}$ because we neglected the $y^{2}$ term from the expansion of 
the Bessel function $I_{0}$ in our computation of the partition function. 
We did so for analytical convenience since otherwise the expressions are lengthy. 
We can improve our estimates of $\lambda_{x}$ in the special case of
$F_{z} = 0$ by performing a simple calculation to show the affect of 
$E_{z}$ on the curve of $\lambda_{x}$ vs. $F_{x}$. When $K_{1} < K_{2}$, 
$F_{z} = 0$ and we do not neglect the $y^{2}$ term from $I_{0}$ the integral 
for the partition function is:
\begin{eqnarray}
 Z_{1} & =& \exp\left(\frac{K_{2}E_{z}^{2}}{2k_{B}T}\right)
 \Bigg(\left[1 + \frac{l^{2}}{4k_{B}^{2}T^{2}}(F_{x}^{2} + F_{y}^{2})\right]
 \int_{-1}^{1} \exp\left(-\frac{(-K_{1}+K_{2})E_{z}^{2}y^{2}}{2k_{B}T}\right) \: \df{y} 
 \nonumber \\
 &  & \qquad \qquad \qquad \quad - \frac{l^{2}}{4k_{B}^{2}T^{2}}(F_{x}^{2} + F_{y}^{2})
 \int_{-1}^{1} \exp\left(-\frac{(-K_{1}+K_{2})E_{z}^{2}y^{2}}{2k_{B}T}\right) y^{2} \: \df{y}\Bigg)
\end{eqnarray}
where we have used $y = \sin\alpha$. Now carrying out the integrals we get
\begin{equation}
\begin{split}
 Z_{1} = \exp\left(\frac{K_{2}E_{z}^{2}}{2k_{B}T}\right)
 \Bigg(&\left[1 + \frac{l^{2}}{4k_{B}^{2}T^{2}}(F_{x}^{2} + F_{y}^{2})\right]
 \frac{1}{E_{z}}\sqrt{\frac{2\pi k_{B}T}{K_{2}-K_{1}}}
 \erf\left(\frac{\sqrt{K_{2}-K_{1}}E_{z}}{\sqrt{2k_{B}T}}\right) \\
 &- \frac{l^{2}}{4k_{B}^{2}T^{2}}(F_{x}^{2} + F_{y}^{2})
 \bigg[\frac{1}{E_{z}}\sqrt{\frac{2\pi k_{B}T}{K_{2}-K_{1}}}
 \frac{k_{B}T}{(K_{2}-K_{1})E_{z}^{2}}
 \erf\left(\frac{\sqrt{K_{2}-K_{1}}E_{z}}{\sqrt{2k_{B}T}}\right) \\
 &- \exp\left(-\frac{(K_{2}-K_{1})E_{z}^{2}}{2k_{B}T}\right)
    \frac{2k_{B}T}{(K_{2}-K_{1})E_{z}^{2}}\bigg]\Bigg)
\end{split}
\end{equation}
So, the free energy is
\begin{eqnarray}
 G(\mathbf{F},\mathbf{E},T) & = -\frac{K_{2}E_{z}^{2}}{2}
 -k_{B}T\log\Bigg(&\left[1 + \frac{l^{2}}{4k_{B}^{2}T^{2}}(F_{x}^{2} + F_{y}^{2})\right]
 \frac{1}{E_{z}}\sqrt{\frac{2\pi k_{B}T}{K_{2}-K_{1}}}
 \erf\left(\frac{\sqrt{K_{2}-K_{1}}E_{z}}{\sqrt{2k_{B}T}}\right) \nonumber \\
 &  & - \frac{l^{2}}{4k_{B}^{2}T^{2}}(F_{x}^{2} + F_{y}^{2})
 \bigg[\frac{1}{E_{z}}\sqrt{\frac{2\pi k_{B}T}{K_{2}-K_{1}}}
 \frac{k_{B}T}{(K_{2}-K_{1})E_{z}^{2}}
 \erf\left(\frac{\sqrt{K_{2}-K_{1}}E_{z}}{\sqrt{2k_{B}T}}\right) \nonumber \\
 &  & \qquad - \exp\left(-\frac{(K_{2}-K_{1})E_{z}^{2}}{2k_{B}T}\right)
    \frac{2k_{B}T}{(K_{2}-K_{1})E_{z}^{2}}\bigg]\Bigg) \label{eq:G1aFx}
\end{eqnarray}
The polarization $P_{z} = -\frac{\partial G}{\partial E_{z}}$ and
$\lambda_{x} = -\frac{\partial G}{\partial F_{x}}$. We set
$q = \frac{E_{z}\sqrt{K_{2}-K_{1}}}{\sqrt{2k_{B}T}}$, then the stretch and
polarization are:
\begin{eqnarray}
 &\lambda_{x} = \frac{1}{2}\frac{F_{x}l}{k_{B}T}
 \frac{\frac{\sqrt{\pi}}{q}\erf(q) - \frac{\sqrt{\pi}}{2q^{3}}\erf(q)  
 + \frac{1}{q^{2}}\exp(-q^{2})}{[1 + \frac{l^{2}}{4k_{B}^{2}T^{2}}
 (F_{x}^{2} + F_{y}^{2})]\frac{\sqrt{\pi}}{q}\erf(q) 
 -\frac{l^{2}}{4k_{B}^{2}T^{2}}(F_{x}^{2} + F_{y}^{2})
 (\frac{\sqrt{\pi}}{2q^{3}}\erf(q) - \frac{\exp(-q^{2})}{q^{2}})}, 
 \label{eq:lamxG1aFx} \\
 &P_{z} = K_{2}E_{z} + \frac{\sqrt{k_{B}T(K_{2}-K_{1})}}{\sqrt{2}}
 \frac{(1 + \frac{l^{2}}{4k_{B}^{2}T^{2}}(F_{x}^{2}
 + F_{y}^{2}))(-\frac{\sqrt{\pi}}{q^{2}}\erf(q) + \frac{2}{q}\exp(-q^{2}))}
 {[1 + \frac{l^{2}}{4k_{B}^{2}T^{2}}(F_{x}^{2} + F_{y}^{2})]
 \frac{\sqrt{\pi}}{q}\erf(q) -\frac{l^{2}}{4k_{B}^{2}T^{2}}
 (F_{x}^{2} + F_{y}^{2})(\frac{\sqrt{\pi}}{2q^{3}}\erf(q) 
 - \frac{\exp(-q^{2})}{q^{2}})} \nonumber \\
 &- \frac{\sqrt{k_{B}T(K_{2}-K_{1})}}{\sqrt{2}}
 \frac{\frac{l^{2}}{4k_{B}^{2}T^{2}}(F_{x}^{2} + F_{y}^{2})
 (-\frac{3\sqrt{\pi}}{2q^{4}}\erf(q) + \frac{3}{q^{3}}\exp(-q^{2}) 
 + \frac{2}{q}\exp(-q^{2}))}{[1 + \frac{l^{2}}{4k_{B}^{2}T^{2}}
 (F_{x}^{2} + F_{y}^{2})]\frac{\sqrt{\pi}}{q}\erf(q) 
 -\frac{l^{2}}{4k_{B}^{2}T^{2}}(F_{x}^{2} + F_{y}^{2})(\frac{\sqrt{\pi}}{2q^{3}}
 \erf(q) - \frac{\exp(-q^{2})}{q^{2}})}. \label{eq:PzG1aFx} 
\end{eqnarray}
This shows that $E_{z}$ affects the $\lambda_{x}$ vs. $F_{x}$ curve. The result
of these formulae agrees very well with Monte Carlo simulations 
for $K_{2}>K_{1}$ when $\frac{F_{x}l}{k_{B}T} < 1$ and $E_{z} \leq 2$ for both 
$\lambda_{x}$ and $P_{z}$ vs. $\frac{F_{x}l}{k_{B}T}$ as shown by the black
curves in figure~\ref{fig:fig3EzFx}(a) and (b) for $E_{z} = 2$. Since we have 
expanded the Bessel function $I_{0}$ assuming small 
$\frac{l\sqrt{F_{x}^{2} + F_{y}^{2}}}{k_{B}T}$ we do 
not expect the above approximation to hold for large values of $F_{x}$ (results
not shown). To examine how these formulae perform at high $E_{z}$ we refer
the reader to the green curves in figure~\ref{fig:fig3EzFx}(c) and (d).
Panel (c) shows good agreement of eqn. (\ref{eq:lamxG1aFx}) with Monte Carlo 
simulations only for $\frac{F_{x}l}{k_{B}T} \leq 1$ when $E_{z} = 10$. Panel 
(d) shows that the polarization predicted by eqn. (\ref{eq:PzG1aFx}) agrees 
with Monte Carlo simulations for small forces; however, since the 
polarization does not change much for larger forces eqn. (\ref{eq:PzG1aFx}) 
is predictive for larger $F_{x}$ too. 

When $K_{2} < K_{1}$ and $F_{z} = 0$ and $F_{x} \neq 0$ we can once again show the effect of 
$E_{z}$ on the $\lambda_{x}$ vs. $F_{x}$ curve by computing a partition
function as follows: 
\begin{eqnarray}
 Z_{1} & =& \exp\left(\frac{K_{2}E_{z}^{2}}{2k_{B}T}\right)
 \bigg(\left[1 + \frac{l^{2}}{4k_{B}^{2}T^{2}}(F_{x}^{2} + F_{y}^{2})\right]
 \int_{-1}^{1} \exp\left(-\frac{(-K_{1}+K_{2})E_{z}^{2}y^{2}}{2k_{B}T}\right) \: \df{y} 
 \nonumber \\
 &  & - \frac{l^{2}}{4k_{B}^{2}T^{2}}(F_{x}^{2} + F_{y}^{2})
 \int_{-1}^{1} \exp\left(-\frac{(-K_{1}+K_{2})E_{z}^{2}y^{2}}{2k_{B}T}\right) y^{2} \: \df{y}\bigg)
\end{eqnarray}
where this time $K_{1} > K_{2}$, so that the argument of the exponentials above
is positive. Now carrying out the integrals we get
\begin{eqnarray}
 Z_{1} & = \exp\left(\frac{K_{2}E_{z}^{2}}{2k_{B}T}\right)
 \Bigg(&\left[1 + \frac{l^{2}}{4k_{B}^{2}T^{2}}(F_{x}^{2} + F_{y}^{2})\right]
 \frac{1}{E_{z}}\sqrt{\frac{2\pi k_{B}T}{K_{1}-K_{2}}}
 \erfi\left(\frac{\sqrt{K_{1}-K_{2}}E_{z}}{\sqrt{2k_{B}T}}\right) \nonumber \\
 &  & - \frac{l^{2}}{4k_{B}^{2}T^{2}}(F_{x}^{2} + F_{y}^{2})
 \bigg[-\frac{1}{E_{z}}\sqrt{\frac{2\pi k_{B}T}{K_{1}-K_{2}}}
 \frac{k_{B}T}{(K_{1}-K_{2})E_{z}^{2}}
 \erfi\left(\frac{\sqrt{K_{1}-K_{2}}E_{z}}{\sqrt{2k_{B}T}}\right) \nonumber \\
 &  & + \exp\left(\frac{(K_{1}-K_{2})E_{z}^{2}}{2k_{B}T}\right)
    \frac{2k_{B}T}{(K_{1}-K_{2})E_{z}^{2}}\bigg]\Bigg)
\end{eqnarray}
So, the free energy is
\begin{eqnarray}
 G(\mathbf{F},\mathbf{E},T) & = -\frac{K_{2}E_{z}^{2}}{2}
 -k_{B}T\log\Bigg(&\left[1 + \frac{l^{2}}{4k_{B}^{2}T^{2}}(F_{x}^{2} + F_{y}^{2})\right]
 \frac{1}{E_{z}}\sqrt{\frac{2\pi k_{B}T}{K_{1}-K_{2}}}
 \erfi\left(\frac{\sqrt{K_{1}-K_{2}}E_{z}}{\sqrt{2k_{B}T}}\right) \nonumber \\
 &  & - \frac{l^{2}}{4k_{B}^{2}T^{2}}(F_{x}^{2} + F_{y}^{2})
 \bigg[-\frac{1}{E_{z}}\sqrt{\frac{2\pi k_{B}T}{K_{1}-K_{2}}}
 \frac{k_{B}T}{(K_{1}-K_{2})E_{z}^{2}}
 \erfi\left(\frac{\sqrt{K_{1}-K_{2}}E_{z}}{\sqrt{2k_{B}T}}\right) \nonumber \\
 &  & + \exp\left(\frac{(K_{1}-K_{2})E_{z}^{2}}{2k_{B}T}\right)
    \frac{2k_{B}T}{(K_{1}-K_{2})E_{z}^{2}}\bigg]\Bigg) \label{eq:G1bFx}
\end{eqnarray}
The polarization $P_{z} = -\frac{\partial G}{\partial E_{z}}$ and
$\lambda_{x} = -\frac{\partial G}{\partial F_{x}}$. We set
$q = \frac{E_{z}\sqrt{K_{1}-K_{2}}}{\sqrt{2k_{B}T}}$, then the stretch and
polarization are:
\begin{eqnarray}
 &\lambda_{x} = \frac{1}{2}\frac{F_{x}l}{k_{B}T}
 \frac{\frac{\sqrt{\pi}}{q}\erfi(q) + \frac{\sqrt{\pi}}{2q^{3}}\erfi(q)  
 - \frac{1}{q^{2}}\exp(q^{2})}{[1 + \frac{l^{2}}{4k_{B}^{2}T^{2}}
 (F_{x}^{2} + F_{y}^{2})]\frac{\sqrt{\pi}}{q}\erfi(q) 
 -\frac{l^{2}}{4k_{B}^{2}T^{2}}(F_{x}^{2} + F_{y}^{2})
 (-\frac{\sqrt{\pi}}{2q^{3}}\erfi(q) + \frac{\exp(q^{2})}{q^{2}})},  
 \label{eq:lamxG1bFx} \\
 &P_{z} = K_{2}E_{z} + \frac{\sqrt{k_{B}T(K_{1}-K_{2})}}{\sqrt{2}}
 \frac{(1 + \frac{l^{2}}{4k_{B}^{2}T^{2}}(F_{x}^{2}
 + F_{y}^{2}))(-\frac{\sqrt{\pi}}{q^{2}}\erfi(q) + \frac{2}{q}\exp(q^{2}))}
 {[1 + \frac{l^{2}}{4k_{B}^{2}T^{2}}(F_{x}^{2} + F_{y}^{2})]
 \frac{\sqrt{\pi}}{q}\erfi(q) -\frac{l^{2}}{4k_{B}^{2}T^{2}}
 (F_{x}^{2} + F_{y}^{2})(-\frac{\sqrt{\pi}}{2q^{3}}\erfi(q) 
 + \frac{\exp(q^{2})}{q^{2}})} \nonumber \\
 &- \frac{\sqrt{k_{B}T(K_{1}-K_{2})}}{\sqrt{2}}
 \frac{\frac{l^{2}}{4k_{B}^{2}T^{2}}(F_{x}^{2} + F_{y}^{2})
 (\frac{3\sqrt{\pi}}{2q^{4}}\erfi(q) - \frac{3}{q^{3}}\exp(q^{2}) 
 + \frac{2}{q}\exp(q^{2}))}{[1 + \frac{l^{2}}{4k_{B}^{2}T^{2}}
 (F_{x}^{2} + F_{y}^{2})]\frac{\sqrt{\pi}}{q}\erfi(q) 
 -\frac{l^{2}}{4k_{B}^{2}T^{2}}(F_{x}^{2} + F_{y}^{2})
 (-\frac{\sqrt{\pi}}{2q^{3}}\erfi(q) + \frac{\exp(q^{2})}{q^{2}})}.
 \label{eq:PzG1bFx}
\end{eqnarray}
This shows that $E_{z}$ affects the $\lambda_{x}$ vs. $F_{x}$ curve. The result
of these formulae agrees very well with Monte Carlo simulations 
for $K_{1}>K_{2}$ when $\frac{F_{x}l}{k_{B}T} < 1$ and $E_{z} \leq 2$ for both 
$\lambda_{x}$ and $P_{z}$ vs. $\frac{F_{x}l}{k_{B}T}$ as shown by the magenta 
curves in figure~\ref{fig:fig3EzFx}(a) and (b) for $E_{z} = 2$. Since we have 
expanded the Bessel function $I_{0}$ assuming small 
$\frac{l\sqrt{F_{x}^{2} + F_{y}^{2}}}{k_{B}T}$ we do 
not expect the above approximation to hold for large values of $F_{x}$ (results
not shown). Nonetheless, at high $E_{z}$ they perform well over a broad range 
of $\frac{F_{x}l}{k_{B}T}$ as seen from the magenta curves in 
figure~\ref{fig:fig3EzFx}(c) and (d) for $E_{z} = 10$. 

\subsection{Large force}
\begin{figure}
\centering
\includegraphics{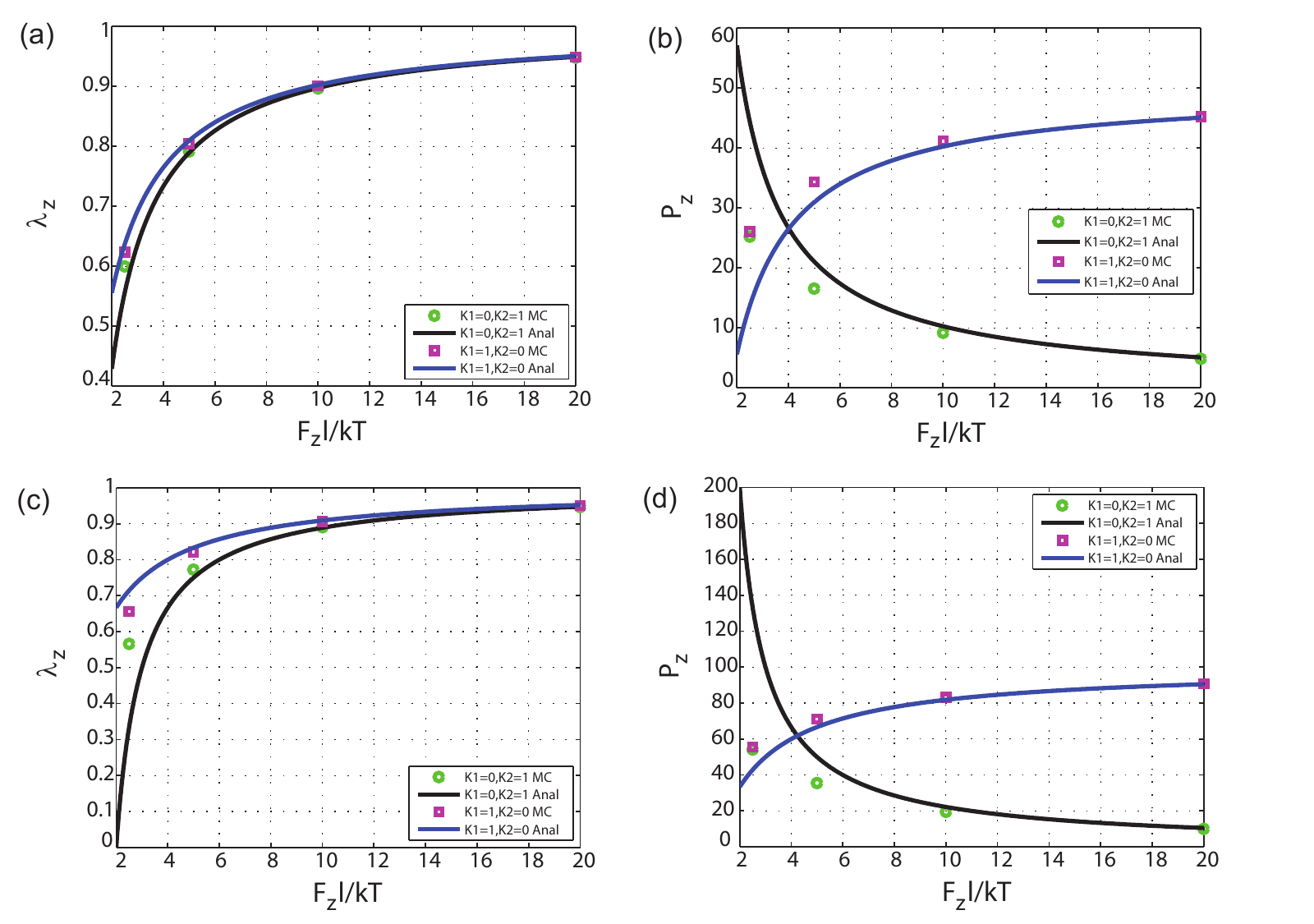}
\caption{Stretch $\lambda_{z}$ and polarization $P_{z}$ for large values of
$F_{z}$ obtained from differentiating the free energy in 
eqn. (\ref{eq:GlargeFz}). The curve for $\lambda_{z}$ is obtained from 
eqn. (\ref{eq:lamzGlargeFz}) and that for $P_{z}$ from 
eqn. (\ref{eq:PzGlargeFz}). Panels (a) and (b) correspond to $E_{z} = 0.5$ 
and panels (c) and (d) to $E_{z} = 1$. We take $N = 100$, $k_{B}T = 1$ and 
examine both $K_{1} = 0, K_{2} = 1$ and $K_{1} = 1, K_{2} = 0$.} 
\label{fig:fig6largeFz}
\end{figure}

\paragraph{Case 2a:} When the force $\mathbf{F}$ is very large then it could be
more convenient to assume that the $z$-axis is aligned with the force, so the 
potential energy of the applied force on a link is
\begin{equation}
 V_{mech} = -Fl\cos\theta,
\end{equation}
where $l$ is the length of the link. The total energy which will enter the
partition sum for a single link is 
\begin{equation} \label{eq:enrgcase1}
\begin{split}
 V(\theta,\phi) = V_{elec} + V_{mech}   
 =& -\frac{K_{1} - K_{2}}{2}
 \left(E_{x}\sin\theta\cos\phi + E_{y}\sin\theta\sin\phi + E_{z}\cos\theta\right)^{2} \\
 &- \frac{K_{2}}{2}\left(E_{x}^{2} + E_{y}^{2} + E_{z}^{2}\right) - Fl\cos\theta.
\end{split}
\end{equation}
Consider the case when $F$ is very large so that $V_{mech}$ is much larger than
$V_{elec}$. Then, for fixed $\phi$, $\exp(-V(\theta,\phi))$ will be largest 
near $\theta = 0$ and decrease as $\theta \rightarrow \pi$. So, we will
approximate $\cos\theta \approx 1 - \frac{\theta^{2}}{2}$ and 
$\sin\theta \approx \theta$ near $\theta = 0$. We will now re-write 
$V(\theta,\phi)$ under the assumption that terms higher order than $\theta^{2}$
can be neglected. Then, the energy takes the following form
\begin{equation}
 V(\theta,\phi) = A(\phi)\theta^{2} + B(\phi)\theta + C(\phi),
\end{equation}
where
\begin{eqnarray}
 A & =&  \frac{Fl}{2} - \frac{K_{1}-K_{2}}{2}\left[E_{x}\cos^{2}\phi 
 + E_{y}\sin^{2}\phi + 2E_{x}E_{y}\sin\phi\cos\phi\right] 
 + \frac{(K_{1} - K_{2})E_{z}^{2}}{2}, \\  
 B & =& -(K_{1} - K_{2})\left(E_{x}E_{z}\cos\phi + E_{y}E_{z}\sin\phi\right), \\
 C & =& -Fl - \frac{K_{2}}{2}(E_{x}^{2} + E_{y}^{2}) - \frac{K_{1}}{2}E_{z}^{2}.
\end{eqnarray} 
Now we can integrate out $\theta$ in the partition sum 
eqn. (\ref{eq:Z1parti}) using Gaussian integrals as follows.
\begin{eqnarray}
 Z_{1} & =& \int_{0}^{2\pi} \int_{0}^{\pi}\exp\left(-
 \frac{A(\phi)\theta^{2} + B(\phi)\theta + C(\phi)}{k_{B}T}\right) \theta \:
 \df{\theta} \df{\phi} \nonumber \\  
 & \approx & \int_{0}^{2\pi} \int_{0}^{\infty}\exp\left(-
 \frac{A(\phi)\theta^{2} + B(\phi)\theta + C(\phi)}{k_{B}T}\right) \theta \:
 \df{\theta} \df{\phi} \nonumber \\  
 & =&\int_{0}^{2\pi}\frac{1}{4A^{3/2}(\phi)}\exp\left(-\frac{C}{k_{B}T}\right)\left(
 2k_{B}T\sqrt{A(\phi)} - \sqrt{\pi k_{B}T}B(\phi)
 \exp\left(\frac{B^{2}(\phi)}{4A(\phi)k_{B}T}\right)
 \erfc\left(\frac{B(\phi)}{2\sqrt{A(\phi)k_{B}T}}\right)\right) \: \df{\phi} \nonumber
\end{eqnarray}  
The last integral over $\phi$ is a one dimensional integral, but it is 
difficult and must be done numerically. If we assume $B(\phi) \ll A(\phi)$, 
because $F$ is large, then the second term in the parentheses in the integrand
will be negligible in comparison to the first. In that case $Z_{1}$ simplifies
to 
\begin{eqnarray}
 Z_{1} & =& \frac{k_{B}T}{2}\exp\left(-\frac{C}{k_{B}T}\right)
 \int_{0}^{2\pi} \frac{\df{\phi}}{\frac{Fl}{2} 
 - \frac{K_{1}-K_{2}}{2}(E_{x}\cos\phi + E_{y}\sin\phi)^{2} 
 + \frac{(K_{1} - K_{2})E_{z}^{2}}{2}} \nonumber \\  
 & =& \frac{k_{B}T}{2}\exp\left(-\frac{C}{k_{B}T}\right)\frac{2\pi}{\sqrt{\frac{Fl}{2}
 + \frac{(K_{1}-K_{2})E_{z}^{2}}{2}}\sqrt{\frac{Fl}{2} 
 + \frac{(K_{1}-K_{2})(E_{z}^{2}-E_{x}^{2}-E_{y}^{2})}{2}}}
\end{eqnarray}
The free energy is $G(F,E_{x},E_{y},E_{z},T) = -Nk_{B}T\log Z_{1}$, and it is
given by
\begin{equation} \label{eq:GlargeFz}
\begin{split}
 G(F,E_{x},E_{y},E_{z},T) =& -N\left(Fl + \frac{K_{2}}{2}(E_{x}^{2}
 + E_{y}^{2}) + \frac{K_{1}}{2}E_{z}^{2}\right) -Nk_{B}T\log\pi k_{B} T \\
 & + \frac{Nk_{B}T}{2}\log\left[\frac{Fl}{2} + \frac{(K_{1}-K_{2})E_{z}^{2}}{2}\right]
 + \frac{Nk_{B}T}{2}\log\left[\frac{Fl}{2} + \frac{(K_{1}-K_{2})(E_{z}^{2}
 -E_{x}^{2}-E_{y}^{2})}{2}\right].
\end{split}
\end{equation}
The above expression for $G(F,E_{x},E_{y},E_{z},T)$ is transversely isotropic
with $z$-axis as the symmetry axis. Let us now differentiate it and get
various other quantities. The polarizations are:
\begin{eqnarray}
\label{eq:largeF-Px}
 P_{x} &= -\frac{\partial G}{\partial E_{x}} =& NK_{2}E_{x}
 + Nk_{B}T\frac{(K_{1}-K_{2})E_{x}}{Fl + (K_{1}-K_{2})(E_{z}^{2} - E_{x}^{2} 
  - E_{y}^{2})}, \\
\label{eq:largeF-Py}
 P_{y} &= -\frac{\partial G}{\partial E_{y}} =& NK_{2}E_{y}
 + Nk_{B}T\frac{(K_{1}-K_{2})E_{y}}{Fl + (K_{1}-K_{2})(E_{z}^{2} - E_{x}^{2} 
  - E_{y}^{2})}, \\
\label{eq:largeF-Pz}
 P_{z} &= -\frac{\partial G}{\partial E_{z}} =& NK_{1}E_{z}
 - Nk_{B}T\frac{(K_{1}-K_{2})E_{z}}{Fl + (K_{1}-K_{2})E_{z}^{2}}
 \nonumber \\ & &- Nk_{B}T\frac{(K_{1}-K_{2})E_{z}}{Fl + (K_{1}-K_{2})(E_{z}^{2} - E_{x}^{2} 
  - E_{y}^{2})}. \label{eq:PzGlargeFz}
\end{eqnarray} 
The end-to-end distance is:
\begin{equation}
 \langle z \rangle = -\frac{\partial G}{\partial F} = Nl 
  - \frac{Nlk_{B}T}{4}\left[
  \frac{1}{\frac{Fl}{2} + \frac{(K_{1}-K_{2})E_{z}^{2}}{2}} +  
  \frac{1}{\frac{Fl}{2} + \frac{(K_{1}-K_{2})(E_{z}^{2}-E_{x}^{2}
   -E_{y}^{2})}{2}}\right].
\end{equation} so that,
\begin{equation} \label{eq:lamzGlargeFz}
 \lambda_{z} = \frac{\langle z\rangle}{L} = 1 -  \frac{k_{B}T}{2}\left[
  \frac{1}{Fl + (K_{1}-K_{2})E_{z}^{2}} +  
  \frac{1}{Fl + (K_{1}-K_{2})(E_{z}^{2}-E_{x}^{2} -E_{y}^{2})}\right],
\end{equation} 
where we denote the stretch $\lambda_{z}$ because the force is $\mathbf{F} = F\mathbf{e}_{z}$. The stretches and polarizations obtained using the above formulae match quite well with data from our Monte Carlo simulations when $F_{z}$ is large and 
$E_{z}$ is small as shown in figure~\ref{fig:fig6largeFz}. 
When $\mathbf{E} = \mathbf{0}$, then we get the familiar result 
$\lambda_{z} = 1 - \frac{k_{B}T}{Fl}$. Recall that for a FJC the force-extension 
relation is $\lambda = \coth\left(Fl / k_{B}T\right) - k_{B}T / Fl 
= \mathcal{L}\left(Fl / k_{B}T\right)$, the Langevin function. 
$\coth(x) \rightarrow 1$ as $x \rightarrow \infty$ in agreement with our result
for large $F_{z}$. Thus, it is not surprising that the force-stretch curve for
small $E_{z}$ is close to the FJC result. 

\begin{figure}
\centering
\includegraphics{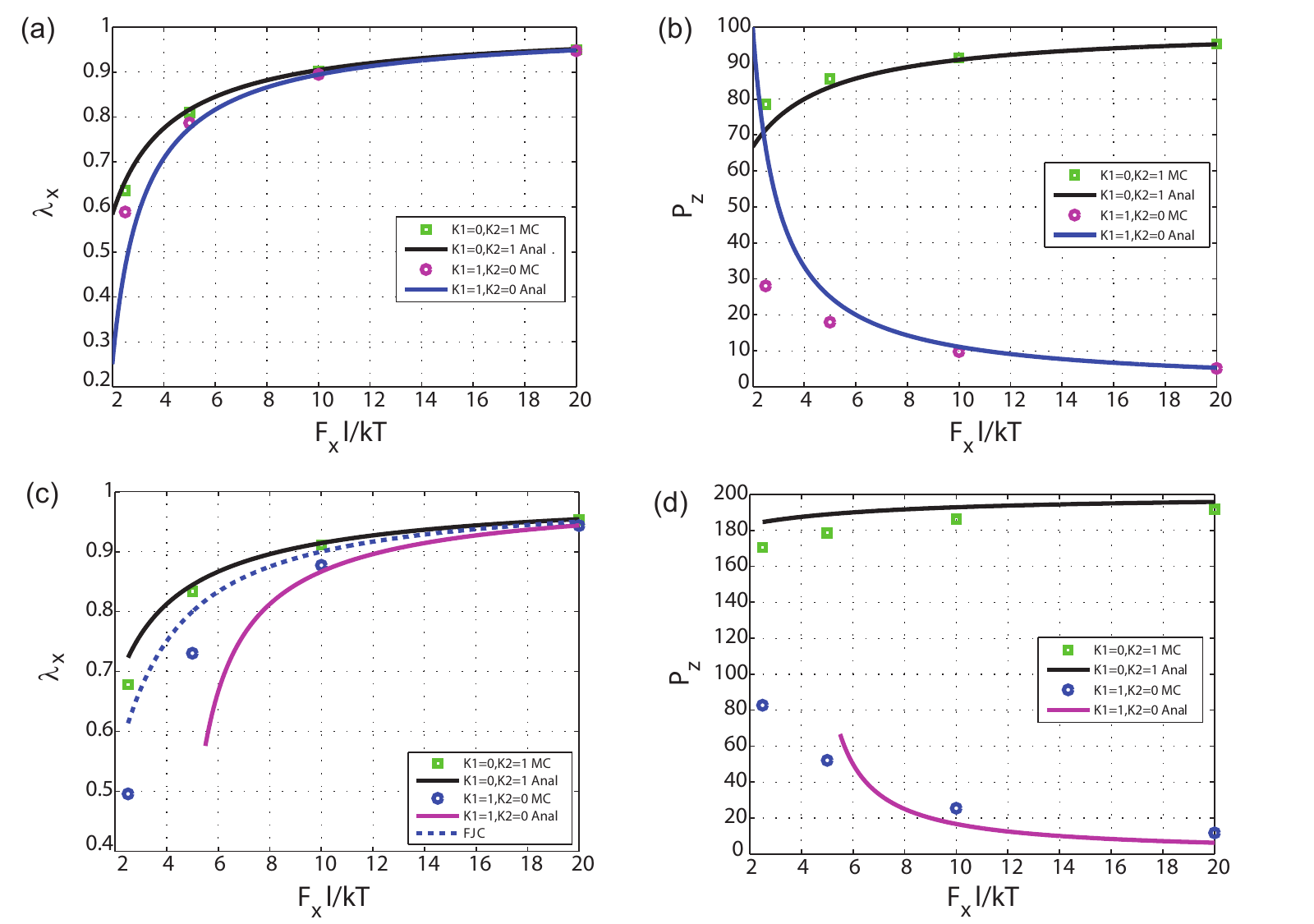}
\caption{Stretch $\lambda_{x}$ and polarization $P_{z}$ for various values of
$F_{x}$ and $E_{z}$ obtained from eqn. (\ref{eq:lamzlargeFx}) and 
eqn. (\ref{eq:PzlargeFx}). Note that $x$ and $z$ are switched in the figure
because it was assumed in the Monte Carlo simulations that the force was along
$x$ and the electric field was along the $z$-direction. We take $N = 100$, $k_{B}T = 1$ 
and consider both cases, 
$K_{1} = 0, K_{2} = 1$ and $K_{1} = 1, K_{2} = 0$. The lines are results of analytical 
formulae. In panels (a) and (b) the symbols are results from Monte 
Carlo simulations with $E_{z} = 1$ and in panels (c) and (d) with $E_{z} = 2$.
The dashed line in panel (c) is the prediction of the FJC formula which assumes
zero electric field.}
\label{fig:fig7EzlargeFx}
\end{figure}

\paragraph{Case 2b:} In the case when the force $\mathbf{F}$ and the electric field $\mathbf{E}$ are perpendicular to each other, say $F_{z} \neq 0, F_{x} = F_{y} = 0$ and $E_{x} \neq 0$, 
$E_{y} = E_{z} = 0$, then 
\begin{eqnarray}
 \lambda_{z} & =& 1 -  \frac{k_{B}T}{2}\left[
  \frac{1}{Fl} +  \frac{1}{Fl - (K_{1}-K_{2})E_{x}^{2}}\right],
 \label{eq:lamzlargeFx} \\
 P_{x} & =& NK_{2}E_{x} 
 + Nk_{B}T\frac{(K_{1}-K_{2})E_{x}}{Fl - (K_{1}-K_{2})E_{x}^{2}}. 
 \label{eq:PzlargeFx}
\end{eqnarray}
In figure~\ref{fig:fig7EzlargeFx} we plot the results of these formulae
with $x$ and $z$ switched because that is how the Monte Carlo simulations were
performed. In other words, the force is along the $x$-direction and the 
electric field is along the $z$-direction. The $\lambda_{x}$ vs. $F_{x}$ and 
$P_{z}$ vs. $F_{x}$ curves match the Monte Carlo data quite well for 
large $F_{x}l/k_{B}T$ $(>6)$ at $E_{z} = 1,2$. As $E_{z}$ becomes large the
estimates from the analytical formulae get worse because these expressions
assume that the electric field is small.

\begin{hlbreakable}
Before moving on, we acknowledge instances in which the analytical and Monte Carlo results exhibit some disparity, and mention the potential of another approach to the statistical mechanics problem considered in this work.
In the absence of dipole-dipole interactions, the integrals for the ensemble averages (e.g. end-to-end vector, net chain dipole, etc.) can be simplified to integration in 2 dimensions.
As a result, numerical quadrature is a tenable choice for performing the integration.
In particular, the trapezoidal rule has been shown to converge rapidly for the integration of periodic, analytic functions~\cite{trefethen2014exponentially}.
In \fref{app:trap}, we generate limited results of ensemble averages using the trapezoidal rule as a means for 1) briefly probing its suitability for statistical mechanics problems similar to the one considered in this work and 2) as a means of determining which approach--the analytical or Monte Carlo--may be more or less correct in some of the few instances in which there are disparity in their results.
\end{hlbreakable}

\subsection{Special case: Aligned electric field and force} \label{sec:special}
Before moving on, we consider the special case of both the electric field and force being aligned, as it allows us to evaluate the partition function exactly. Once again, we take the $z$-direction along the electric field. Then, the partition function takes the simplified form:
\begin{equation}
  Z_1 = 2 \pi \exp\left(\frac{E^2 K_2}{2 k_B T}\right) \int_0^\pi \exp\left(-\frac{\frac{1}{2} E^2 \left(K_2 - K_1\right) \cos^2 \theta - F l \cos \theta}{k T}\right) \sin \theta \: \df{\theta}
\end{equation}
which is axisymmetric and, hence, can be evaluated exactly:
\begin{equation}
  Z_1 = \pi^{3/2} \exp\left(\frac{E^2 K_2}{2 k_B T} + \frac{F^2 l^2}{4 \EDimless k_B T}\right) \sqrt{\frac{k_B T}{\EDimless}} \left(\erf\left(\frac{2 \EDimless + F l}{2 \sqrt{\EDimless k_B T}}\right) + \erf\left(\frac{2 \EDimless - F l}{2 \sqrt{\EDimless k_B T}}\right)\right)
\end{equation}
where, for convenience, we have made the definition $\EDimless \coloneqq E^2 \left(K_2 - K_1\right) / 2$.
Next, the chain stretch in the $z$-direction is given by
\begin{equation} \label{eq:special-stretch}
  \lambda = \frac{F l}{2 \EDimless} + \frac{
    \exp\left(-\frac{\left(2 \EDimless + F l\right)^2}{4 \EDimless k_B T}\right)\left(\exp\left(\frac{2 F l}{k_B T}\right) - 1\right)
    }{
    \erf\left(\frac{F l - 2 \EDimless}{2 \sqrt{\EDimless k_B T}}\right) - \erf\left(\frac{F l + 2 \EDimless}{2 \sqrt{\EDimless k_B T}}\right)
}\sqrt{\frac{k_B T}{\pi \EDimless}}.
\end{equation}
We can obtain the compliance of the chain by taking $\partial \lambda / \partial F$.
However, this expression is quite complicated.
To probe the chain compliance in the limit of small force~\footnote{One may wonder why it was necessary to evaluate the partition function exactly if we eventually intended to expand its solution in a small parameter. 
However, as seen in previous works on dielectric polymer chains~\cite{grasingerIPtorque}, there are important differences between using an approximate formulation before ensemble averaging and making approximations after ensemble averaging.}, we perform a Taylor expansion of \eqref{eq:special-stretch} about $F = 0$ and obtain:
\begin{equation}
  \lambda = \frac{F l}{2 \EDimless} \left(1 - \frac{
      2\sqrt{\EDimless}\exp\left(-\frac{\EDimless}{k_B T}\right)
  }{
  \sqrt{\pi k_B T} \erf\left(\sqrt{\frac{\EDimless}{k_B T}}\right)
}\right) + \mathcal{O}\left(F^2\right)
\end{equation}
Neglecting higher order terms (i.e. $\mathcal{O}\left(F^2\right)$), we can easily obtain the small force compliance:
\begin{equation}
  c = \frac{\partial \lambda}{\partial F} = \frac{l}{2 \EDimless} \left(1 - \frac{
      2\sqrt{\EDimless}\exp\left(-\frac{\EDimless}{k_B T}\right)
  }{
  \sqrt{\pi k_B T} \erf\left(\sqrt{\frac{\EDimless}{k_B T}}\right)
}\right).
\end{equation}
We now consider the compliance in various limits:
\begin{enumerate}
  \item When $\EDimless \rightarrow 0$, we recover the (small force) compliance for the classical FJC: \begin{equation}
      c \rightarrow \frac{l}{3 k_B T}.
    \end{equation}
The compliance diverges as $k_B T \rightarrow 0$ which is characteristic of a second-order phase transition~\cite{ericksen1998introduction}.
For the Ising model, the order parameter is magnetism (i.e. average magnetic moment).
When $k_B T \rightarrow 0$, all the magnetic dipoles want to be aligned either up or down.
A magnetism of zero is an unstable equilibrium when the applied magnetic field is zero.
There needs to be a small perturbation in the applied magnetic field, either up or down, for the system to magnetize.
In this case, the magnetic susceptibility is diverging.
Similarly, for the classical FJC, $\lambda = 0$ is an unstable equilibrium when $F = 0$.
We need a small perturbation in the force and then we have a jump, in that direction, to a fully stretched chain.
It is also useful to consider why collective behavior occurs in either case.
For the Ising model, collective behavior emerges because magnetic dipoles interact with their neighbors.
For the FJC, collective behavior emerges from a different kind of interaction;
instead, collective behavior occurs because tension is transmitted through the chain for any finite force.
In contrast to a nearest neighbor interaction, it is a kind of global, mechanical interaction.
  \item Interestingly, when $\EDimless \rightarrow -\infty$ (i.e. $K_1 > K_2$ and $E \rightarrow \infty$), the compliance takes the form
    \begin{equation}
      c \rightarrow \frac{l}{k_B T}.
    \end{equation}
    We see here that again the compliance diverges, but it does so 3 times more rapidly than in the $\EDimless \rightarrow 0$ limit.
    This has a nice geometric interpretation to it.
    In the limit of $\EDimless \rightarrow -\infty$, the electrical potential energy landscape has infinitely deep wells at $\hat{\mathbf{n}} = \hat{E}$ and $\hat{\mathbf{n}} = -\hat{E}$, so that, effectively, the links only have a single degree of freedom.
    They can either be in the field direction or opposite the field direction.
    The dimensionality of the system, in a certain sense, has been reduced from 3D to 1D; hence the compliance diverging 3 times more rapidly as $k_B T \rightarrow 0$.
  \item For any finite $\EDimless > 0$ (i.e. $K_2 > K_1$ and $E \neq 0$), the compliance takes the form
    \begin{equation}
      c \rightarrow \frac{l}{2}
    \end{equation}
in the limit $k_B T \rightarrow 0$.
Thus, in this case, a phase transition does not occur, even in the limit of zero temperature.
This can be understood by considering that, when $K_2 > K_1$, links prefer to be orthogonal to the field direction which, by assumption, is also the direction of the applied force.
So although the chain no longer has an entropic stiffness, there is a stiffness induced by the electric field which keeps the compliance from diverging.
We remark that, while the case of $F$ orthogonal to $E$ and $K_2 > K_1$ cannot be investigated by exactly evaluating the partition function, the results of this section suggest that this case would also exhibit a second-order phase transition where, if $\EDimless \rightarrow \infty$, the compliance diverges as $l / 2 k_B T$.
\end{enumerate}

The link flipping algorithm outlined in \Fref{sec:flipping} performs well even as the system approaches $\EDimless \rightarrow -\infty$.
However, we suspect that the MCMC simulations exhibited poor convergence when $K_1 > K_2$, the electric field was large, the applied force was large, and the electric field and force were orthogonal, or nearly orthogonal, to each other, because of a kind of symmetry-breaking that occurs as the magnitude of the force increases.
It is possible that this is a sign of the emergence of a first-order phase transition; however, analyzing this case presents a formidable challenge and is left for future work.
It is interesting that phase transitions occur at all, and illustrate the complexity of the thermodynamics of dielectric polymer chains, given that we have neglected dipole-dipole interactions.
The influence of dipole-dipole interactions is also a topic the authors intend to explore in future work.

\section{Thermodynamic limit and comparison of with end-to-end vector ensemble} \label{sec:thermo}

One of the motivations in investigating the fixed force ensemble for dielectric polymer chains was to make a comparison with previous results in the fixed end-to-end vector ensemble so as to probe the existence and qualities of the thermodynamic limit; \hl{that is, the limit in which the differences in behavior between the two ensembles becomes negligible}.
\hl{For polymer chains in the absence of pairwise interactions or interactions with external fields (e.g. electromagnetic fields), the differences between the end-to-end vector and force ensembles are negligible in the long chain limit~\cite{weiner2012statistical,treloar1975physics}--which~\citet{treloar1975physics} shows to be $N \geq 100$.
It is not clear, however, whether the differences between the ensembles at $N \geq 100$ are still negligible when the monomers of the chains are interacting with an electric field.}
Unfortunately, the equations for the statistical mechanics of the dielectric polymer chain make it difficult, if not impossible, to evaluate the partition function exactly in either ensemble; and, further, different approximations are more or less appropriate in either case.
It is difficult then to determine--when there are differences between the two ensembles--whether or not the differences are physically meaningful or a consequence of choice of approximation.
For these reasons, a general quantitative comparison between the two ensembles is beyond reach.
However, we can be confident that when there are notable phenomena that the two ensembles share that they are physically relevant and a qualitative feature of dielectric chains in the thermodynamic limit.
Along these lines, we compare the two ensembles \hl{at $N = 100$} and note the following similarities:
\begin{enumerate}
    \item The free energy is a transversely isotropic function of its arguments.
    For the fixed force ensemble, the free energy depends on $\bfE \cdot \bfE$, $\bfE \cdot \bfF$, and $\bfF \cdot \bfF$.
    Similarly, for the fixed end-to-end vector ensemble, the free energy is a function of $\bfE \cdot \bfE$, $\bfE \cdot \bfr$, and $\bfr \cdot \bfr$~\cite{grasinger2020statistical}.
    \item The polymer stiffness depends on $|\bfE|$ and the direction of the applied force relative to the electric field in the force ensemble; and its stiffness depends on $|\bfE|$, and the direction of end-to-end vector relative to the electric field in the end-to-end ensemble.
    For $K_2 > K_1$, chains become more stiff with increasing $|\bfE|$ when forced or stretched in the field direction (see \Fref{fig:fig1K1lK2} or \Fref{fig:ensemble-comparison}).
    For $K_1 > K_2$, chains become less stiff with increasing $|\bfE|$ when forced or stretched in the field direction (see \Fref{fig:fig2K2lK1} or \Fref{fig:ensemble-comparison}).
    This occurs because, for $K_2 > K_1$, when $\hat{\bfn}$ is in the direction of the electric field, the link is in a (local) maximum electrical energy state; and, for $K_1 > K_2$, when $\hat{\bfn}$ is in the direction of the electric field, the link is in a (local) minimum electrical energy state.
    Therefore, an electric-field-induced symmetry breaking of the (otherwise isotropic) chain stiffness is a feature of dielectric chains regardless of ensemble choice and, consequently, a feature of the thermodynamic limit.
    For dielectric elastomer networks, this symmetry breaking is an important component for intrinsic electrostriction~\cite{grasingerIPtorque}.
   
    Beyond qualitative agreement, \Fref{fig:ensemble-comparison} shows an excellent quantitative agreement between the two ensembles regarding the force-stretch relationship for $|\bfE| \sqrt{|K_1 - K_2| / k_B T} = \frac{1}{2}, 1,$ and $2$; and for both cases: $K_1 < K_2$ (a) and $K_1 > K_2$ (b).
    We remark that the agreement is poor however when $K_1 > K_2$ and $|\bfE| \sqrt{|K_1 - K_2| / k_B T} \ll 1$, as the approximate expression for the force ensemble can suffer from numerical instabilities in this regime.
    Similarly, the agreement is poor when $K_1 > K_2$ and $|\bfE| \sqrt{|K_1 - K_2| / k_B T} \gtrsim 5$, as the approximate expression for the end-to-end vector ensemble begins to produce nonphysical results in this regime~\cite{grasinger2020statistical}.
    \begin{figure}
        \centering
        \includegraphics[width=\linewidth]{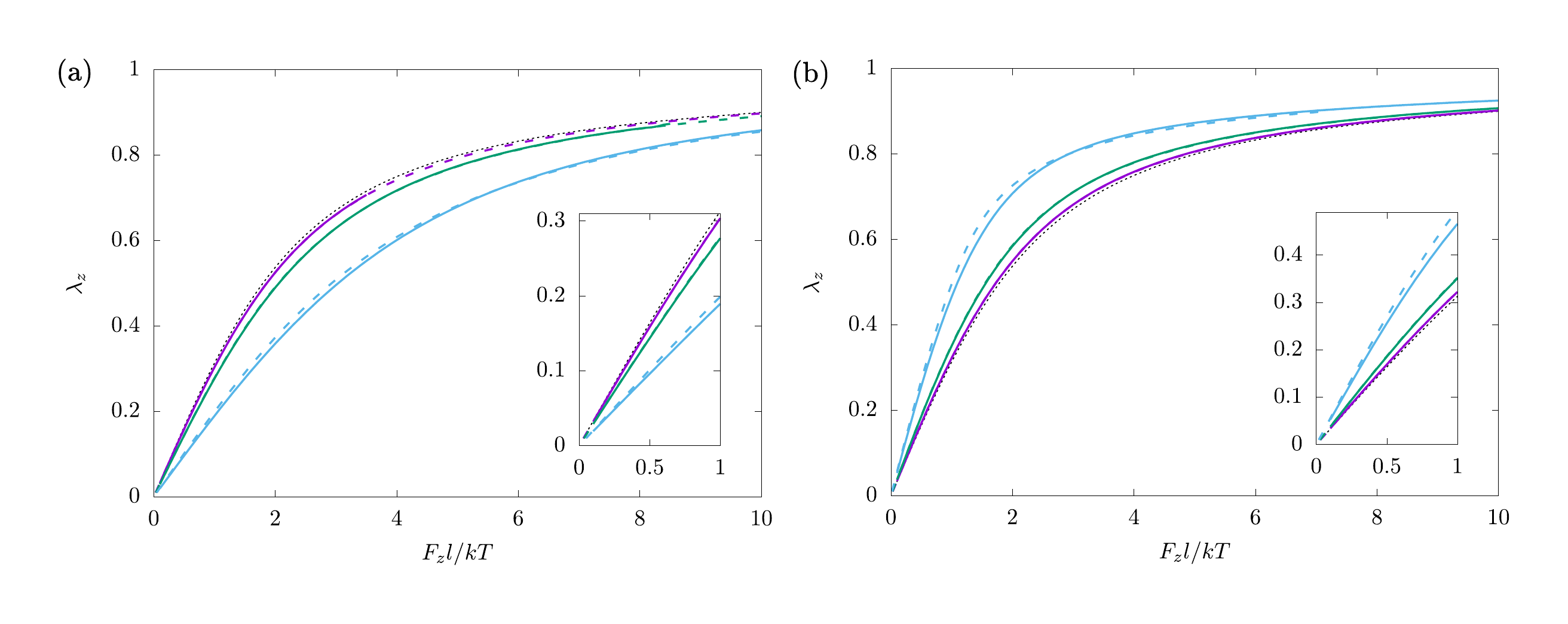}
        \caption{Comparison of the force-stretch relationship as obtained by the force ensemble (solid lines) and end-to-end ensemble~\cite{grasinger2020statistical} (dashed lines) for \hl{$N = 100$, and} $|\bfE| \sqrt{|K_1 - K_2| / k_B T} = \frac{1}{2}, 1,$ and $2$ (magenta, green, and blue lines, respectively; the black dotted line shows the classical FJC result).
        (a) shows the $K_1 < K_2$ case and (b) shows the $K_1 > K_2$ case.}
        \label{fig:ensemble-comparison}
    \end{figure}
    \item The chain polarization depends on the applied force (end-to-end vector) (see \Fref{fig:fig1K1lK2}.b and \Fref{fig:fig2K2lK1}.b) in a nontrivial way because as the force (stretch) increases, the average monomer alignment is shifted to the direction of force (stretch) and the polarization of a link depends on $\hat{\bfn}$.
    Importantly, the force (and end-to-end vector) dependence of the chain polarization is such that the chain polarization may be misaligned with the electric field (see \eqref{eq:largeF-Px}--\eqref{eq:largeF-Pz}).
    For an individual chain, the misalignment of its polarization with the electric field means that it will experience an electrostatic torque~\cite{grasinger2020statistical}.
    For an elastomer network, the misalignment can propagate to the continuum-scale and lead to asymmetric electromechanical modes of actuation~\cite{grasingerIPtorque}.
\end{enumerate}

\section{Conclusions}
\label{sec:conclusions}

The primary accomplishment of this work is to provide new insights into the coupled electromechanical behavior of polymers.
In contrast to top-down approaches that are based on developing continuum models based on experimental observations, our approach starts from discrete models at the scale of monomers and uses statistical mechanics to coarse-grain to the continuum scale.
An important advance in this work is that it goes beyond the mean-field approximation that was used in prior approaches, e.g. \cite{cohen2016electroelasticity,grasinger2020statistical}.
This was enabled through the use of a fixed chain force ensemble, rather than the dual fixed chain extension ensemble that was used in prior approaches.
In addition, to compare the closed-form approximate expressions with numerical approaches, we develop a new adaptation of Monte Carlo (MC) methods inspired by ferromagnetic Ising model approaches.
We highlight that all our closed-form predictions agree well with the MC simulations in the appropriate limits.
Further, our results agree with the fixed chain extension ensemble in various qualitative aspects, as well as in those cases where quantitative comparison can be made.

Based on the new analysis and numerical strategies, we derive expressions for the stretch and polarization, that simplify to closed-form in the following regimes:
\begin{enumerate}
    \item \textit{The component of the force perpendicular to the field is small.}
    We consider 
    the cases $K_1<K_2$ and $K_2<K_1$, with $K_1$ and $K_2$ the polarizability along and normal, respectively, to the monomer.
    We find that when $K_1 < K_2$, the chain stretches less in response to a force (or is less responsive to force) than a freely-jointed chain with zero electric field.  When $K_2 < K_1$, the chain stretches more in response to a force (or is more responsive to force) than a freely-jointed-chain. This can be rationalized by noticing that the electric field tends to align the monomers normal to itself when $K_1<K_2$, whereas the field tends to align the monomers parallel to itself when $K_2 < K_1$. In both settings, we obtain closed-form expressions in the limit that an (appropriately nondimensionalized) electric field is much larger than the applied force.
    
    \item \textit{The electric field and the force are perpendicular.}
    This setting corresponds to the practically-important case of dielectric elastomers used in a typical actuation mode.
    We find that in this setting, chains composed of monomers with $K_2 < K_1$ are more responsive compared to monomers with $K_1 < K_2$.
    The reason is that both the electric field and the force work together to align the links normal to the field \cite{cohen2016electromechanical}.
    
    \item \textit{The force is dominant over the electric field.}
    We find that $K_2 < K_1$ leads to a chain that is more responsive than the FJC case, and $K_1 < K_2$ leads to the opposite.

    \item \textit{The electric field and the force are aligned.}
    In this setting, we derive an exact expression for the stretch and an approximation for its stiffness. 
    We show that if $K_1>K_2$ the chain stiffness diverges when $k_B T \rightarrow 0$. 
    We further demonstrate that if $K_2>K_1$, the chain is stable under all loads. This analysis hints at the possibility that when $K_2>K_1$, the stiffness diverges when the electric field is perpendicular to the force.
\end{enumerate}

We finally highlight important limitations of our work, and the potential for future work to go beyond these limitations.
\begin{enumerate}
    \item Our final results are not posed in terms of a standard free energy that can be directly used in continuum calculations.
    That is, we do not obtain an explicit (Helmholtz) free energy density as a function of the deformation gradient and polarization, that could then be used in solving boundary value problems at the continuum scale.
    Instead, our energy densities have as arguments the electric field and the force.
    In principle, these energy densities have the same information as the Helmholtz free energy density, and are dual to each other in a standard Legendre transform.
    However, performing a Legendre transform to obtain closed-form expressions is challenging, and an important goal for the future.

    \item We neglect dipole-dipole interactions, following prior work \cite{cohen2016electroelasticity,grasinger2020statistical}.
    While this is justifiable in regimes in which the external applied field is large, it can potentially miss important physics.
    In particular, the dipole-dipole interactions make the problem highly nonlocal, which is challenging for analysis but could lead to rich physics.
    
    \item The use of a fixed force ensemble has fewer kinematic constraints than the typical fixed chain extension ensemble.
    Therefore, it provides the possibility of better capturing potential instabilities and phase transformations.
    Exploring such instabilities could provide new strategies for nonlinear enhancement of the electromechanical coupling.

\end{enumerate}

\section*{Software availability}

A version of the code developed for this work is available at  \\ \url{https://github.com/grasingerm/polymer-stats}.

\begin{acknowledgments}

PKP acknowledges partial support for this work from an NSF grant NSF CMMI 1662101. 
K.D. acknowledges support from NSF (1635407), ARO (W911NF-17-1-0084), ONR (N00014-18-1-2528), BSF (2018183), and AFOSR (MURI FA9550-18-1-0095). GdB acknowledges support from BSF (2018183).
We acknowledge NSF for computing resources provided by Pittsburgh Supercomputing Center.
\end{acknowledgments}

\appendix

\section{Trapezoidal rule} \label{app:trap}

\begin{hlbreakable}
In the absence of dipole-dipole interactions, the integrals for the ensemble averages (e.g. end-to-end vector, net chain dipole, etc.) can be simplified to integration in 2 dimensions (see, for example, \eqref{eq:Z1parti}--\eqref{eq:polarization}).
Since the dimensionality of the integration space is not too large, numerical quadrature is a tenable choice for performing the integration.
The trapezoidal rule is a kind of numerical quadrature which exhibits rapid convergence~\cite{trefethen2014exponentially} for integrating functions which are both analytic and periodic; thus, it represents a promising approach for \eqref{eq:Z1parti}--\eqref{eq:polarization}.

Consider $I = \int_a^b f(x) \: \df{x}$.
The trapezoidal rule consists of discretizing the interval into $m + 1$ points, $x_0 = a < x_1 < x_2 < \dots < x_{m-1} < b = x_m$, and approximating the area under $f$ by summing the area of the $m$ trapezoids with the vertices $\left\{ \left(x_{i-1}, 0\right), \left(x_{i-1}, f\left(x_{i-1}\right)\right), \left(x_{i}, 0\right), \left(x_{i}, f\left(x_{i}\right)\right) \right\}, i = 1, \dots, m$.
For computational simplicity (and efficiency), we assume that the points are evenly spaced and let $h \coloneqq \left(b - a\right) / m$ such that $h = x_i - x_{i-1}$ for all $i$.
Then,
\begin{equation}
	I \approx I_t = h \left(\frac{f(a) + f(b)}{2} + \sum_{i=1}^{m-1} f(x_i)\right)
\end{equation}
and $\left|I_t - I\right| \rightarrow 0$ as $m \rightarrow \infty$.
%Remarkably, when $f$ is analytic and periodic over $\left[a, b\right]$, then
%\begin{equation}
%	\left|I_m - I\right| \leq \frac{\left(b - a\right) \sup_{x \in \left[a, b\right]} \left|f(x)\right|}{}
%\end{equation}
An alternative way to understand the trapezoidal rule is that it corresponds with approximating $f$ by a piecewise linear function %(which agree with $f$ at the points $x_0, x_1, \dots, x_m$) 
and then integrating the approximate function over $\left[a, b\right]$.
In general, the error decays as $\mathcal{O}\left(h^2\right)$ but for analytic, periodic functions the method convergences \emph{exponentially}~\cite{trefethen2014exponentially}.

Here we test the trapezoidal rule for approximating ensemble averages by generating results for some of the limited cases in which the analytical and Monte Carlo results showed some disparity.
In \fref{fig:fig8fig5withtrap}, the results for $\lambda_z$ (panel (a)) and $P_z$ (panel (b)) as a function of $F_z l / k T$ are shown for the analytical approach (solid line), the trapezoidal rule (dashed lines), and the Monte Carlo simulations (symbols).
As expected, the trapezoidal rule agrees quite well, in general, which highlights its suitability as a method for statistical mechanics problems of this type.
Where there is some disparity between the analytical and Monte Carlo results, the trapezoidal rule appears to agree more closely to the Monte Carlo simulations at small $F_z$ (i.e. $F_z l / k T \leq 0.5$) and more closely to the analytical results for $F_z l / k T > 0.5$.
This suggests that the noise that appears in the Monte Carlo results when the force becomes significant ($F_z l / k T > 0.5$) is indeed nonphysical.
A similar effect can be seen in \fref{fig:fig9fig2withtrap} which shows the analytical, Monte Carlo, and trapezoidal rule results for $\lambda_{z}$ as a function of $E_z$ for fixed $F_{z} l / k T = 0.5$ and $1.0$.
There, the Monte Carlo data appears noisy relative to the analytical and trapezoidal rule approaches when both the force and electric field are significant ($E_z \geq 5$).

\begin{figure}[hbt!]
	\centering
	\includegraphics{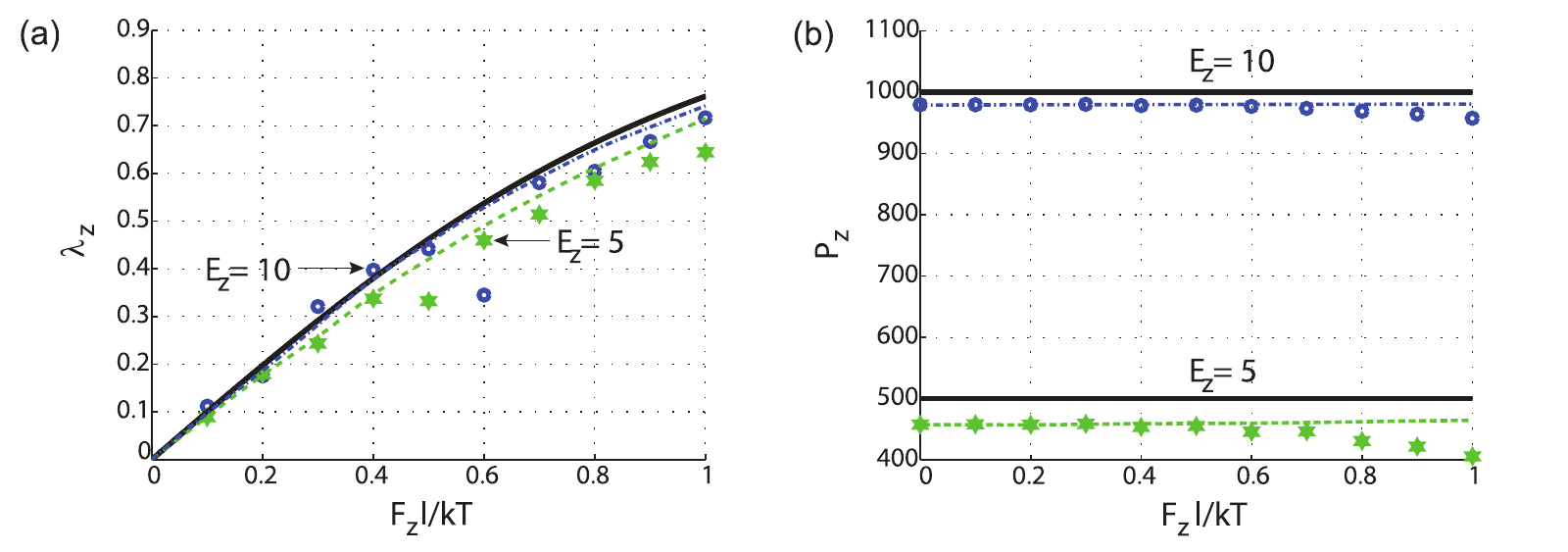}
	\caption{\hl{(a) Stretch, $\lambda_{z}$, and (b) polarization, $P_{z}$, for various values of
		$F_{z}$ and $E_{z}$. We assume $N=100$, $K_{1}=1$, $K_{2}=0$ and $k_{B}T=1$. 
		The solid lines are results of analytical formulae eqn. (\ref{eq:lamztanh})
		and eqn. (\ref{eq:Pztanh}); the dashed lines are the results of the trapezoidal rule 
		with $E_z = 5$ as green dashed and $E_z = 10$ as blue dash-dot; and the symbols are results 
		from Monte Carlo simulations with $E_{z} = 5$ as green stars, and $E_{z} = 10$ as blue circles.}} 
	\label{fig:fig8fig5withtrap}
\end{figure}

\begin{figure}[hbt!]
	\centering
	\includegraphics{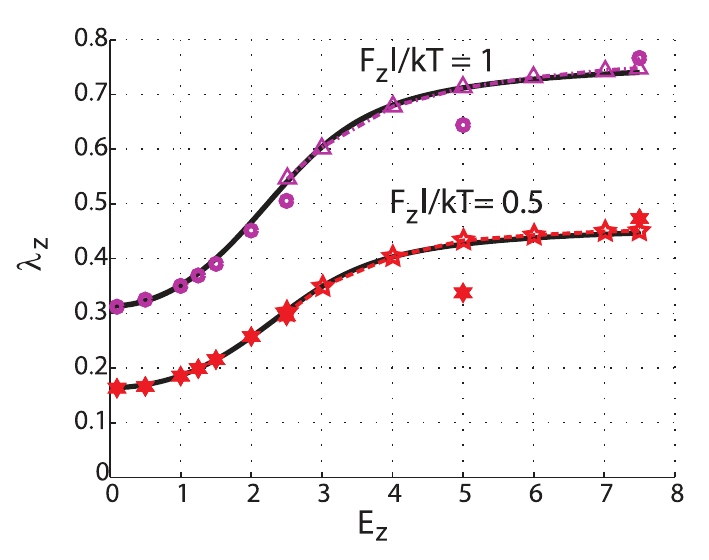}
	\caption{\hl{Stretch, $\lambda_{z}$, as a function of $E_z$ for fixed
		$F_{z} l / k T = 0.5$ and $1.0$. We assume $N=100$, $K_{1}=1$, and $K_{2}=0$. 
		The solid lines are results of analytical formulae; the dashed lines with points are the results of the trapezoidal rule 
		with $F_z l / k T = 0.5$ as red dashed with open stars and $F_z l / k T = 1.0$ as magenta dash-dot with triangles; and the symbols are results 
		from Monte Carlo simulations with $F_z l / k T = 0.5$ as red stars, and $F_z l / k T = 1.0$ as magenta circles.}} 
	\label{fig:fig9fig2withtrap}
\end{figure}

In summary, numerical quadrature, and in particular, the trapezoidal rule present an alternative approach to the computational statistical mechanics of dielectric polymer chains in the absence of dipole-dipole interactions.
Given the noise of the Monte Carlo results when the effects of both the force and electric field are significant, quadrature rules can be seen as the more robust computational approach for the system considered in this work.
However, it is also important to consider the following potential limitations:
\begin{enumerate}
	\item While the convergence of the trapezoidal rule is rapid for the types of integrals considered in this work, it may still be slow relative to Monte Carlo simulations when the probability distribution has a single sharp peak (i.e. only one of $E_z$ or $F_z$ is significant).
	In such cases, only a small subset of $\pSpace$ contributes meaningfully to the integral.
	The importance sampling of the MCMC method focuses computational resources primarily at the peak of the probability distribution which is contributing significantly to the integral, whereas the trapezoidal rule (at least the variant considered in this appendix) samples uniformally over the interval, often making potentially costly computations that result in negligible contributions to the final result.
	There are, however, variants of the trapezoidal rule which adaptively use a higher density of integration points near localized regions with sharp features such as peaks and a coarser set of points elsewhere~\cite{van1976adaptive,berntsen1991adaptive}.
	These adaptive variants of the trapezoidal rule would likely have comparable (or perhaps even greater) efficiency than MCMC importance sampling--even when the probability distribution has a single sharp peak.
	Yet another possible approach is that of $p$-adaptive quadratures such as the Clenshaw-Curtis rule~\cite{clenshaw1960pcubature} which was used to study dielectric polymer chains in~\cite{grasinger2020statistical}.
	A more detailed comparison of these various methods is outside the scope of this work.
	\item More importantly, quadrature rules are no longer tenable when considering dielectric polymer chains with dipole-dipole interactions as the integrals for ensemble averages are over a high dimensional space, $2 N$.
	Since dielectric polymers with dipole-dipole interactions are a topic of interest for future work, we have primarily restricted our attention towards developing MCMC techniques.
\end{enumerate}

\end{hlbreakable}

\bibliographystyle{unsrtnat}
\bibliography{master}

\end{document}